\documentclass[number,preprint,3p]{elsarticle}

\usepackage{color}
\usepackage{soul}
\usepackage{graphicx}
\usepackage{subcaption}
\usepackage{algorithm}
\usepackage{bm}
\usepackage[colorlinks]{hyperref}% Add hyper-ref for equations, figures, etc.
\makeatletter
\gdef\urlauthor#1#2{\g@addto@macro\@elsuads{\let\corref\@gobble%
     \def\@@tmp{#1}\raggedright\eadsep
     {\ttfamily\url{\expandafter\strip@prefix\meaning\@@tmp}}\space(#2)%
     \def\eadsep{\unskip,\space}}%
}
\gdef\emailauthor#1#2{\stepcounter{ead}%
     \g@addto@macro\@elseads{\raggedright%
      \let\corref\@gobble\def\@@tmp{#1}%
      \eadsep{\ttfamily\href{mailto:\expandafter\strip@prefix\meaning\@@tmp}{\expandafter\strip@prefix\meaning\@@tmp}}
      (#2)\def\eadsep{\unskip,\space}}%
}
\makeatother
\usepackage{amssymb}
\usepackage{amsthm}
\usepackage{amsmath}
\usepackage{amssymb}
\usepackage{mathtools}
\usepackage{dsfont}
\usepackage{booktabs}
\usepackage{url}
\usepackage{scrextend}
\usepackage{epstopdf}
\usepackage{float}
\usepackage{tcolorbox}
\usepackage{tablefootnote}
\usepackage[perpage]{footmisc}% Renew the footnote numbering for each page
\usepackage{bbm}
\usepackage{lineno}

\def\r{\mathbb{R}}
\def\rn{\mathbb{R}^n}
\def\defi{\coloneqq}
\def\tr{^\intercal}

\newcommand{\Prob}[1]{\mathbb{P}\left(#1\right)}

\newcommand{\1}[2]{\mathbb{I}_{#1}\left(#2\right)}
\newcommand{\vect}[1]{\boldsymbol{#1}}


\biboptions{numbers,sort&compress,square}
\journal{arXiv}
\begin{document}
%\linenumbers
	\begin{frontmatter}
		\renewcommand{\thefootnote}{\fnsymbol{footnote}}
		\title{A physics and data co-driven surrogate modeling method \\ for high-dimensional rare event simulation}
		\author[1]{Jianhua Xian}
		\author[1]{Ziqi Wang\corref{cor1}}
         \ead{ziqiwang@berkeley.edu}
         \cortext[cor1]{Corresponding author}
		\address[1]{Department of Civil and Environmental Engineering, University of California, Berkeley, United States}
		\begin{abstract}
			This paper presents a physics and data co-driven surrogate modeling method for efficient rare event simulation of civil and mechanical systems with high-dimensional input uncertainties. The method fuses interpretable low-fidelity physical models with data-driven error corrections. The hypothesis is that a well-designed and well-trained simplified physical model can preserve salient features of the original model, while data-fitting techniques can fill the remaining gaps between the surrogate and original model predictions. The coupled physics-data-driven surrogate model is adaptively trained using active learning, aiming to achieve a high correlation and small bias between the surrogate and original model responses in the critical parametric region of a rare event. A final importance sampling step is introduced to correct the surrogate model-based probability estimations. Static and dynamic problems with input uncertainties modeled by random field and stochastic process are studied to demonstrate the proposed method.		
		\end{abstract}
		
		\begin{keyword}
		surrogate modeling \sep active learning \sep importance sampling \sep high-dimensional \sep rare event simulation\sep uncertainty quantification
		\end{keyword}
		
	\end{frontmatter}
	
	\renewcommand{\thefootnote}{\fnsymbol{footnote}}
	
	%% main text
	\section{Introduction}
	
	\noindent Uncertainty quantification (UQ) aims at quantifying and understanding the influence of ubiquitous uncertainties arising in science and engineering. Rare event simulation is a challenging branch of UQ with numerous engineering applications, such as the reliability of aerospace systems \cite{morio2015estimation}, resilience of critical infrastructures \cite{zio2021risk}, and safety of nuclear power plants \cite{jiang2021triso}. Sampling and surrogate modeling-based methods are two general approaches for rare event simulation, although other more specialized techniques exist \cite{dang2020mixture,xu2022adaptive,li2009stochastic,chen2019direct,xian2021seismic,lyu2022unified}.  Monte Carlo simulation \cite{rubinstein1998modern,landau2015guide} is typically insensitive to the dimensionality and nonlinearity of computational models. However, for rare event simulation, even the advanced variance-reduction techniques \cite{au2001estimation,kurtz2013cross,wang2016cross,engel2023bayesian,grigoriu2020data,papaioannou2016sequential,wang2019hamiltonian,xian2024relaxation} would require thousands of samples, restricting their applications to expensive computational models. 
	
	Surrogate modeling seeks to replace the original expensive computational model with an efficient substitute. A non-intrusive data-fitting surrogate model is derived directly from training  samples of the input-output pairs of the  computational model. Examples of data-fitting surrogate models include the quadratic response surface \cite{rajashekhar1993new,allaix2011improvement}, {polynomial chaos expansion \cite{sudret2002comparison,marelli2018active,torre2019data}}, support vector machine \cite{hurtado2004examination,roy2023support}, and {Gaussian process (also called Kriging) \cite{kaymaz2005application,bichon2008efficient,echard2011ak,echard2013combined,huang2016assessing,wang2020novel}}, among others. Unlike many regression models, a noise-free Gaussian process is an exact interpolation model, i.e., the predictions at the training points are exact. For non-training points, the Gaussian process offers estimations of prediction variability {\cite{jones2001taxonomy,santner2003design,sudret2012meta}}. This feature facilitates the application of active learning, such that the training set can be adaptively enriched to reduce prediction variability for specified quantities of interest, resulting in improved accuracy and efficiency for Gaussian process-based rare event simulation \cite{bichon2008efficient,echard2011ak}. The active learning-based metamodeling approach can be further combined with advanced sampling techniques \cite{echard2013combined,huang2016assessing}, or adapted to estimate probability distribution functions \cite{wang2020novel}. 
 
 Despite the remarkable success of Gaussian process-based methods in various UQ applications, the model training becomes increasingly challenging, ultimately infeasible, with a growing number of input variables--a phenomenon known as the curse of dimensionality \cite{lataniotis2020extending}. To alleviate this problem, specialized unsupervised \cite{giovanis2018uncertainty,giovanis2020data,dos2022grassmannian,kontolati2022survey} and supervised \cite{jiang2017high,zhou2021active,navaneeth2022surrogate,kim2024adaptive} dimensionality reduction techniques have been proposed to couple with Gaussian process modeling. However, the effectiveness of these techniques is problem-dependent. More critically, many real-world high-dimensional problems do not admit simple low-dimensional representations \cite{jiang2021recursive}. Qualitatively speaking, for a ``high-dimensional input--low-dimensional output" problem, an ``optimal" dimensionality reduction is the computational model itself. It follows that constructing an effective dimensionality reduction can be as challenging as finding an accurate high-dimensional surrogate model. {An alternative approach is to inject domain/problem-specific prior knowledge into the surrogate modeling process. This idea is reflected in the emerging paradigms of scientific machine learning \cite{raissi2019physics,zhu2019physics,linka2022bayesian,meng2023pinn} and multi-fidelity UQ \cite{peherstorfer2018survey,peherstorfer2016multifidelity,kramer2019multifidelity}, enabling them to handle high-dimensional problems.}

	In this paper, we leverage physics-based surrogate modeling to solve high-dimensional rare event estimation problems. This idea is under the broad umbrella of multi-fidelity UQ. %Data-fitting surrogate models are typically versatile but lack interpretability, while physics-based simplified models are readily interpretable but often inaccurate. 
 A physics-based surrogate model can be adapted from the original high-fidelity model in various ad hoc ways, such as domain-specific simplifications \cite{peterson2018overview,han2013improving,held2005gap}, {reduced-order modelings \cite{benner2015survey,kramer2019nonlinear,patsialis2020reduced}}, and relaxations of numerical solvers \cite{peherstorfer2018survey}. In this study, {we address static and dynamic problems with a random field/process as input, typically discretized by 1,000 random variables}. We construct physics-based surrogate models equipped with the properties of (i) {parsimonious}: the surrogate model is parameterized by a few tunable control parameters, and (ii) {compatible with high-dimensional input uncertainties}: the surrogate model can accept high-dimensional input uncertainties either directly as input of the model or indirectly through filtering and coarse-graining operations. For example, the classic equivalent linearization method \cite{crandall2006half,elishakoff2017sixty} for nonlinear random vibration analysis involves constructing linear physical models with random processes as input and output. Therefore, {the} equivalent linearization method can be reformulated into a physics-based surrogate modeling approach for nonlinear random vibration problems \cite{wang2024optimized}. In other engineering applications such as computational fluid dynamics \cite{peterson2018overview}, aerodynamic design \cite{han2013improving}, and climate modeling \cite{held2005gap}, there exist various domain-specific approaches in constructing simplified physics-based models. %General approaches include coarse-graining of spatial and temporal grids and relaxations of tolerances involved in iterative solvers \cite{peherstorfer2018survey}. 
 %To ensure the accuracy of surrogate modeling for rare event simulation, it is necessary to introduce a final importance sampling step to couple the surrogate model simulations with a limited number of the original computational model simulations, known as multi-fidelity importance sampling \cite{peherstorfer2016multifidelity,kramer2019multifidelity}. However, the common practice of multi-fidelity importance sampling relies on the Gaussian mixture model as the importance density, which scales poorly with dimensionality \cite{wang2016cross}. In this work, we adopt a new importance sampling formulation  \cite{wang2022optimized} to address high-dimensional rare event simulations.
	
	Physics-based surrogate models may have inherent errors due to simplifications; therefore, it is promising to introduce a data-driven error correction to fill the gap between the surrogate and original model predictions. {This idea has been investigated recently in \cite{dhulipala2022active,dhulipala2022reliability}, where an active learning-based Gaussian process is trained to correct the low-fidelity model predictions. Inspired by the success of existing works,
 %However, the data-driven surrogates of error corrections therein are constructed on the model inputs directly, and thus the methodology may be restricted to low- and medium-dimensional input uncertainties due to the poor scalability of Gaussian process metamodeling with dimensionality \cite{lataniotis2020extending}. 
 this study is devoted to low probability estimations with high-dimensional input uncertainties. Different from existing works \cite{dhulipala2022active,dhulipala2022reliability}, we construct the data-driven error correction as a function of the output of the physics-based surrogate model, thereby mitigating the curse of dimensionality associated with high-dimensional input uncertainties.} Three critical ingredients--parametric optimization for physics-based surrogate models, heteroscedastic Gaussian process for error corrections, and active learning for effective training--are leveraged to construct and train coupled physics-data-driven surrogate models. Finally, one can apply an importance sampling to couple the surrogate model simulations with limited evaluations of the original model, known as multi-fidelity importance sampling \cite{peherstorfer2016multifidelity,kramer2019multifidelity}. {The common practice of multi-fidelity importance sampling uses parametric distribution models such as the Gaussian mixture as the importance density, which scales poorly with dimensionality \cite{wang2016cross}. In this work, we adopt an importance sampling formulation \cite{wang2024optimized} tailored to high-dimensional rare event simulations.} 
 
 %a critical feature of the present framework targeted for high-dimensionality is that the error correction surrogate is constructed on the 1-dimensional output of the low-fidelity physical model instead of the high-dimensional model input. In essence, this represents a dimensionality reduction for surrogate modeling of error corrections, i.e., the 1-dimensional output space of the low-fidelity physical model can be viewed as a reduced feature space for the high-dimensional input space, which will inevitably result in certain noises for the error corrections. Correspondingly, three main ingredients, i.e., parametric optimization, heteroscedastic Gaussian process, and active learning, are leveraged to fulfill the training of the coupled physics-data-driven surrogate model. Optimization of the parameterized low-fidelity physical model can significantly reduce the noises of error corrections; heteroscedastic Gaussian process can accommodate the noises of error corrections; and active learning can facilitate the adaptive training towards rare event simulation. Finally, an importance sampling adapted for high-dimensional problems \cite{wang2022optimized} is introduced to gurrantee the theoretical correctness of the present approach.
	
	This paper is organized as follows. Section \ref{Sec:surrogatemodel} introduces the general formulation of the coupled physics-data-driven surrogate model. Section \ref{Sec:surrogatemodeltraining} develops a training process for the surrogate model, including optimization of parametric physical models, error corrections using {an} heteroscedastic Gaussian process, and adaptive training by active learning. Section \ref{Sec:ImportanceSampling} introduces an importance sampling scheme to correct the surrogate model-based probability estimations. Section \ref{Sec:Application} demonstrates the performance of the proposed method for high-dimensional rare event simulation problems. {Section \ref{Sec:lim} discusses limitations}. Section \ref{Sec:conclude} provides concluding remarks. The implementation details are provided in \ref{Append:implementationdetails}.
	
	\section{Coupled physics-data-driven surrogate model}\label{Sec:surrogatemodel}
	\noindent Consider an end-to-end computational model $\mathcal{M}:\vect x\in\rn\mapsto y\in\r$ that maps a $n$-dimensional input $\vect x$ into a $1$-dimensional output $y$. The input $\vect x$ is an outcome of a random vector $\vect X$ defined in the probability space $(\rn,\mathcal{B}_n,\mathbb{P}_{\vect X})$, where $\mathcal{B}_n$ is the Borel $\sigma$-algebra on $\rn$, and $\mathbb{P}_{\vect X}$ is the probability measure of $\vect X$ with the distribution function $f_{\vect X}$. The output $y$ is a performance variable such that it defines the rare event of interest through $\{y\leq 0\}$, without loss of generality. Since the source of randomness is from $\vect X$, the rare event probability is typically formulated in the space of $\vect X$, expressed by $\mathbb{P}_{\vect X}(\mathcal{M}(\vect X)\leq 0)$. This problem is challenging in real-world applications, because (i) the dimension of $\vect x$ is high, (ii) the computational model $\mathcal{M}$ involves expensive physics-based simulations, and (iii) the probability to be estimated is small. 
%	\begin{equation}\label{HFmodel}
%		y_{H}=\mathcal{M}_{H}(\vect{x})\,,
%	\end{equation}
%	where $\mathcal{M}_{H}$ is the high-fidelity model that is computationally expensive, $\vect{x}$ is the outcome of a $D$-dimensional vector $\vect{X}$ of basic random variables, and $y_{H}$ is the output quantity of interest. 	
	To reduce the computational cost of $\mathcal{M}$, we construct a simplified (end-to-end) computational model expressed by:
	\begin{equation}\label{LFmodel}
 \begin{aligned}
     &y_{p}=({\mathcal{M}}_p\circ\psi)(\vect x;\vect\theta_p)\,,\\
     &\mathcal{M}_{p}:\vect x'\in\r^{n'}\mapsto y_p\in\r\,,\\
     & \psi:\vect x\in\rn\mapsto\vect x'\in\r^{n'}\,,n'\leq n\,,
 \end{aligned}		
	\end{equation}
	where $\mathcal{M}_{p}$ represents a cheaper physics-based model derived from the original model $\mathcal{M}$, $\psi$ is a filtering or coarse-graining function of the input $\vect x$, ``$\circ$" denotes function composition, and $\vect\theta_p$ are tunable parameters of $\mathcal{M}_{p}$ and/or $\psi$. Depending on the selection of $\mathcal{M}_{p}$, the filtering function $\psi$ can be a trivial identity mapping if $\mathcal{M}_{p}$ and $\mathcal{M}$ share the same input space, or it can be a dimensionality reduction mapping if $\mathcal{M}_{p}$ is defined on a coarser/different input space. However, it is important to notice that ${\mathcal{M}}_p\circ\psi$ and $\mathcal{M}$ share the same source of randomness from $\vect X$, enabling a tuning of ${\mathcal{M}}_p\circ\psi$ (via adjusting $\vect\theta_p$) to maximize the statistical correlation between $Y$ and $Y_p$. 

 To improve the accuracy of Eq.~\eqref{LFmodel}, a data-driven error correction can be introduced, leading to a coupled physics-data-driven surrogate model expressed by:
% and $y_{L}$ is the low-fidelity counterpart of $y_{H}$. The construction of the physics-based surrogate model $\mathcal{M}_{L}$ relies on the domain-specific knowledges and in-depth implementation details \cite{peherstorfer2018survey}. For instance, in the context of nonlinear random vibration, $\mathcal{M}_{H}$ and $\mathcal{M}_{L}$ can be constructed from the nonlinear system and the equivalent linear system, respectively. In a more general setting, one can define $\mathcal{M}_{H}$ and $\mathcal{M}_{L}$ in terms of fine- and coarse-grid spatial and/or temporal discretizations, or high- and low-fidelity nonlinear iterative solvers.	
	%To fill the gap between $y_{H}$ and $y_{L}$, a data-driven surrogate of error corrections can be introduced and constructed on $y_{L}$, leading to a coupled physics-data-driven surrogate model, i.e., 
	\begin{equation}\label{CoupledModel}
		\hat{y}=y_{p}+{\epsilon}(y_{p};\vect\theta_{\epsilon})=(\mathcal{M}_p\circ\psi)(\vect{x};\vect\theta_{p})+({\epsilon}\circ\mathcal{M}_p\circ\psi)(\vect{x};\vect\theta_{p},\vect\theta_{\epsilon})\,,
	\end{equation}
	where ${\epsilon}:y_p\in\r\mapsto\varepsilon\in\r$ is the error correction function constructed using data-fitting methods, and $\vect\theta_{\epsilon}$ are parameters of the data-fitting model. It is worth highlighting that the input of the error correction function is the output of the physics-based surrogate model. This construction is different from existing works \cite{dhulipala2022active,dhulipala2022reliability}, where the error correction term was constructed on the product space of $\vect{x}$ and $\vect x'$. The methodology of \cite{dhulipala2022active,dhulipala2022reliability} may face difficulties for problems with high-dimensional input uncertainties, due to the weak scalability of conventional data-fitting methods such as Gaussian process and polynomial regression \cite{lataniotis2020extending}. In comparison, our construction of the surrogate model can mitigate the curse of dimensionality. However, it comes with a price of having inherent noises in the error correction term, even at the training points of $y_p$. This is because the mapping from $y_p$ to $y$ can be one-to-many. An intuitive restatement is that the physics-based surrogate model cannot capture all details of the original model. To quantify and mitigate the impact of the inherent noises, we use the heteroscedastic Gaussian process \cite{lazaro2013retrieval,rogers2020probabilistic,kim2023estimation} to model ${\epsilon}(y_p)$. This differs from the noise-free and homoscedastic Gaussian process models, as the heteroscedastic Gaussian process model no longer performs exact interpolations and the noise is input-dependent.
		
	\begin{figure}[H]
		\centering
		\includegraphics[scale=0.50]{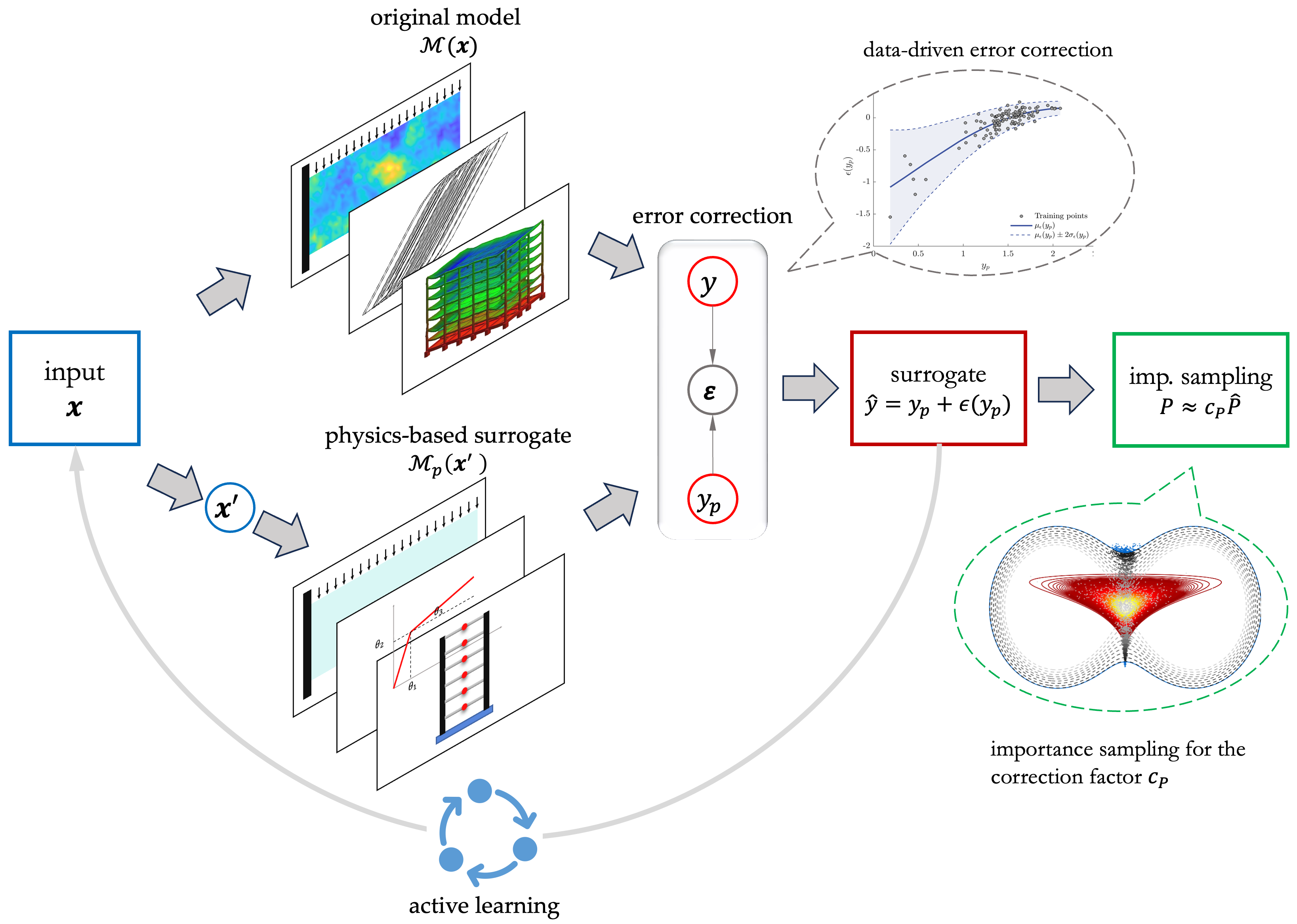}
		\caption{\textbf{Physics and data co-driven surrogate modeling for rare event simulation.} \textit{The physics-based surrogate model $y_p=\mathcal{M}_p(\vect x';\vect\theta_p)$ contains tunable control parameters $\vect\theta_p$ to be optimized in the training process. The error correction function $\epsilon(y_p)$ is modeled in the space of $y_p$ by heteroscedastic Gaussian process to account for input-dependent noises. The active learning adaptively enriches the training set with a learning criterion to highlight contributions from the rare event. The importance sampling estimates the correction factor $c_P$ to improve the probability estimation from the surrogate model}.}
		\label{Fig:Figure0}
	\end{figure}

 Figure \ref{Fig:Figure0} presents a schematic of the proposed surrogate modeling method for rare event simulation. The technical details for optimizing the physics-based surrogate and error correction models are introduced in Sections \ref{Sec:optimization} and \ref{Sec:ErrorCorrection}, respectively. The active learning approach for efficient surrogate model training is described in Section \ref{Sec:ActiveLearning}. Section \ref{Sec:ImportanceSampling} details the importance sampling method for rare event probability estimations.
	
	\section{Training of the coupled physics-data-driven surrogate model }\label{Sec:surrogatemodeltraining}
	\subsection{Optimization of parametric physical models}\label{Sec:optimization}
\begin{figure}[H]
		\centering
		\includegraphics[scale=0.5]{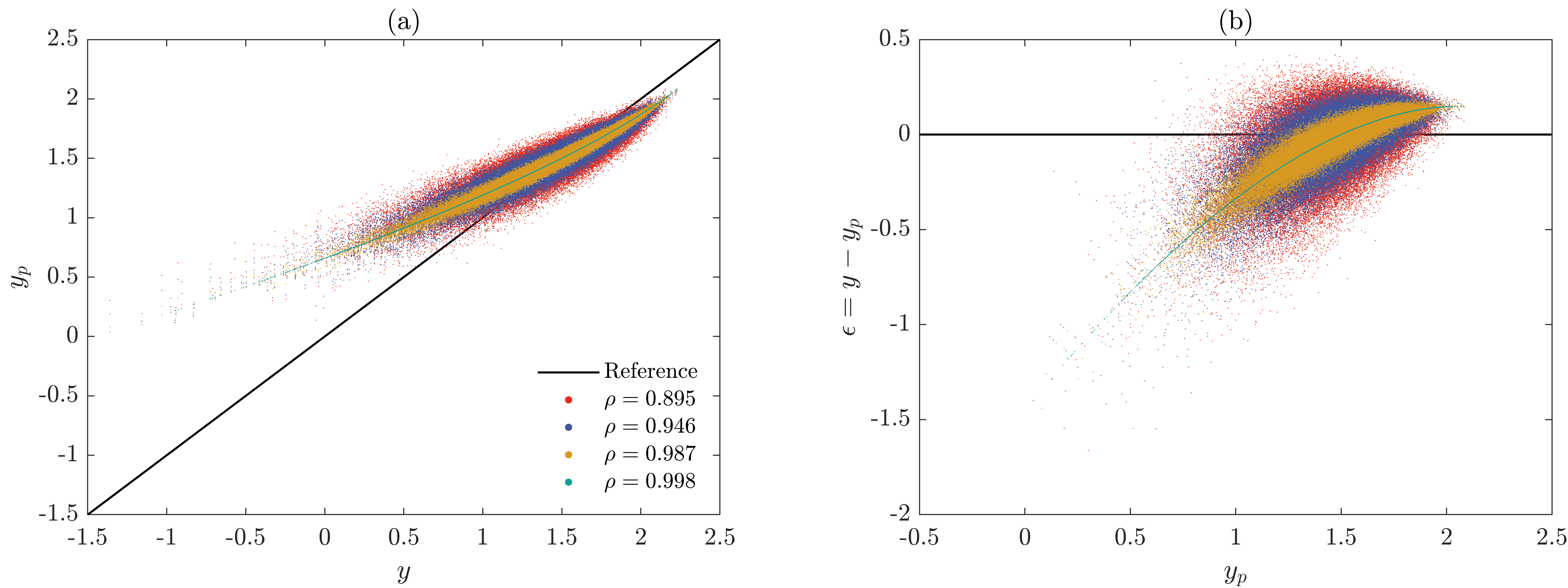}
		\caption{\textbf{{Impact of the correlation between surrogate and original model responses on the noise of the error correction function.}} \textit{For illustration, $Y$ and $Y_{p}$ are assumed to be lognormal with a Pearson correlation coefficient varying among $\rho=0.895$, $0.946$, $0.987$, and $0.998$. It is seen that the noise in the error $\epsilon=y-y_p$ decreases with the increase of the correlation coefficient. It is also clear that an exact interpolation model should be avoided when $\epsilon$ is noisy.}}
		\label{Fig:Figure1}
	\end{figure}

	\noindent In physics-based surrogate modeling, the response predictions of the surrogate model can be inaccurate but still be mildly/highly correlated with that of the original model. The statistical correlation between the surrogate and original model responses has a decisive impact on the noise of the error correction term (see Figure \ref{Fig:Figure1} for a simple illustration)--higher correlation indicates smaller noise. Pearson, Spearman's Rank, and Kendall's Tau correlation coefficients are popular indices for quantifying the correlation between two random variables \cite{hauke2011comparison}. The Pearson correlation coefficient can only reflect the linear correlation, while the Spearman's Rank and Kendall's Tau correlation coefficients can handle the nonlinear dependency. In our context, the physics-based surrogate model, by construction, is a simplification of the original model. Thus, a mildly nonlinear correlation is expected. Consequently, the Pearson correlation coefficient is sufficient. The Spearman's Rank and Kendall's Tau correlation coefficients can be readily implemented into the proposed surrogate modeling method, but we did not observe an improvement in performance for the  numerical examples we have studied. It is worth mentioning that mutual information \cite{taverniers2021mutual,beneddine2023nonlinear} is a more general metric for measuring dependency between two random variables, but the sample estimate of mutual information involves additional assumptions on the joint distributions.

 %A remedy is to consider the correlation between $\1{\leq0}{Y}$ and $\1{\leq0}{Y_p}$, where $\mathbb{I}_{\leq0}:\r\mapsto\{0,1\}$ is a binary indicator function for the rare event. The computational issue of this approach is that the samples generated from $f_{\vect X}(\vect x)$ are unlikely to fall into the rare event and thus the estimation of the Pearson correlation coefficient between $\1{\leq0}{Y}$ and $\1{\leq0}{Y_p}$ would be inaccurate. 
 
 Because rare event probabilities are dominated by conditional distributions, the correlation between $Y$ and $Y_p$ generated by $\vect X\sim f_{\vect X}(\vect x)$ may not be an ideal objective for optimizing the surrogate model. {The issue is that limited samples from $f_{\vect X}(\vect x)$ are unlikely to fall near to the boundary of the rare event, where most of the misclassifications happen. In this work, we adopt active learning to adaptively enrich the training set $\mathcal{D}=\{(\vect x^{i},\mathcal{M}(\vect x^{(i)}))\}$ toward the critical region around $\{\mathcal{M}(\vect x)=0\}$. Using the training set $\mathcal{D}$, the sample correlation coefficient emphasizes the contribution from the rare event. This is essentially equivalent to redefining the correlation between $\mathcal{M}(\vect X)$ and $(\mathcal{M}_p\circ\psi)(\vect X)$ using an active learning-controlled importance sampling distribution rather than the original $f_{\vect X}(\vect x)$.} To summarize, the optimization for the physics-based surrogate model $(\mathcal{M}_{p}\circ\psi)(\vect x;\vect\theta_p)$ is formulated as:
	\begin{equation}\label{OptimizationModel}
 \begin{aligned}
     \vect{\theta} ^{\ast }_{p}&=\mathop{\arg\max}\limits_{\vect{\theta}_{p}}\rho_{YY_p}(\vect{\theta}_{p}|\mathcal{D})\\
     &=\mathop{\arg\max}\limits_{\vect{\theta}_{p}}\frac{\left \langle (\mathcal{M}(\vect{X})- \langle \mathcal{M}(\vect{X})\rangle _{\mathcal{D}})\left((\mathcal{M}_{p}\circ\psi)(\vect{X};\vect{\theta}_{p})- \langle (\mathcal{M}_{p}\circ\psi)(\vect{X};\vect{\theta}_{p})\rangle _{\mathcal{D}}\right)\right \rangle _{\mathcal{D}}}{\sqrt{\left \langle (\mathcal{M}(\vect{X})- \langle \mathcal{M}(\vect{X})\rangle _{\mathcal{D}})^2\right \rangle _{\mathcal{D}} \left \langle ((\mathcal{M}_{p}\circ\psi)(\vect{X};\vect{\theta}_{p})- \left \langle (\mathcal{M}_{p}\circ\psi)(\vect{X};\vect{\theta}_{p}) \right \rangle _{\mathcal{D}})^2\right \rangle _{\mathcal{D}}}}\,,
 \end{aligned}
	\end{equation}
 where $\langle\cdot\rangle$ denotes sample mean. 
	%where $\rho(\vect{\theta}_{L})$ is the correlation coefficient between the high-fidelity model response $Y_{H}=\mathcal{M}_{H}(\vect{X})$ and the low-fidelity parameterized physical model response $Y_{L}(\vect{\theta}_{L})=\mathcal{M}_{L}(\vect{X};\vect{\theta}_{L})$, i.e.,
	%\begin{equation}\label{correlationcoefficient}
		%\rho(\vect{\theta}_{L})=\frac{\mathbb{C}ov_f[Y_{H},Y_{L}(\vect{\theta}_{L})]}{\sqrt{\mathbb{V}ar_f[Y_{H}]\mathbb{V}ar_f[Y_{L}(\vect{\theta}_{L})]}}\,,
	%\end{equation}
   % where $\mathbb{C}ov_f$ and $\mathbb{V}ar_f$ denote respectively the covariance and variance with respect to the probability density function of $\vect{X}$, i.e., $f_{\vect{X}}(\vect{x})$, and $\vect{\theta}_{L}$ denotes a vector containing the parameters of the low-fidelity physical model.
	
	%Notably, Eq.\eqref{OptimizationModel} can be interpreted as finding a low-fidelity physical model that is most positively correlated with the original high-fidelity model. In fact, it is also desirable to derive a low-fidelity physical model that is most negatively correlated with the original high-fidelity model (i.e., maximizing $-\rho$ instead in Eq.\eqref{OptimizationModel}). However, such a physical model generally does not have practical significance and therefore is not considered herein.
 
 Before solving Eq.~\eqref{OptimizationModel}, we need to design $\vect\theta_p$ for the physics-based surrogate model, i.e., determine which parameters are tunable. For specific applications,  engineering judgement can be used to design $\vect\theta_p$. A more objective approach is to adopt a parsimonious principle to select a minimum number of parameters that can achieve a relatively high $\max_{\vect\theta_p}\rho_{YY_p}(\vect{\theta}_{p}|\mathcal{D})$. This principle can be materialized as an incremental approach to start with a single tunable parameter and iteratively augment if $\max_{\vect\theta_p}\rho_{YY_p}(\vect{\theta}_{p}|\mathcal{D})$ can be significantly improved. In the simulation test cases of civil and mechanical systems, we have investigated the use of elastic modulus, damping ratio, stiffness, and yield displacement as tunable parameters, where $\max_{\vect\theta_p}\rho_{YY_p}(\vect{\theta}_{p}|\mathcal{D})$ can typically achieve $0.95$. 

 	Finally, the optimization in Eq.~\eqref{OptimizationModel} can be solved by gradient-free metaheuristic algorithms \cite{conn2009introduction}. Provided with a training set $\mathcal{D}$, solving Eq.~\eqref{OptimizationModel} does not involve additional simulations of the original model, but it requires simulations of the physics-based surrogate model. If $\mathcal{D}$ is enriched by active learning, Eq.~\eqref{OptimizationModel} needs to be re-solved; the solution from the previous training set can be used as a warm initial guess.

	\subsection{Error corrections using heteroscedastic Gaussian process}\label{Sec:ErrorCorrection}
	\noindent Given a training set $\mathcal{D}=\{(\vect x^{i},\mathcal{M}(\vect x^{(i)}))\}$ and the solution of Eq.~\eqref{OptimizationModel}, we obtain the training set for the error correction, $\mathcal{D}_\epsilon=\{(y_p^{(i)},\varepsilon^{(i)})\}$, where $y_p^{(i)}=(\mathcal{M}_p\circ\psi)(\vect x^{(i)};\vect\theta_p^*)$ and $\varepsilon^{(i)}=\mathcal{M}(\vect x^{(i)})-(\mathcal{M}_p\circ\psi)(\vect x^{(i)};\vect\theta_p^*)$. Using $\mathcal{D}_\epsilon$, we train a heteroscedastic Gaussian process: \cite{lazaro2013retrieval,rogers2020probabilistic,kim2023estimation} to model $\epsilon(y_p)$, expressed by:
% As it is almost impossible to seek a low-fidelity physical model that is completely correlated with the original high-fidelity model, certain noises inevitably exist in the error corrections and should be incorporated into the data-driven surrogate modeling. For this purpose, a heteroscedastic Gaussian process \cite{lazaro2013retrieval,rogers2020probabilistic,kim2023estimation} that can account for input-dependent noises is adopted herein to model the noisy error corrections. Suppose the error correction term $\hat{\epsilon}(y_{L})$ in Eq.\eqref{CoupledModel} can be considered as a realization of a heteroscedastic Gaussian process, i.e.,
	\begin{equation}\label{hGPerrorcorrection}
		{\epsilon}(y_{p};\vect\theta_\epsilon)=f(y_{p};\vect\theta_f)+\tau(y_{p};\vect\theta_\tau)\,,
	\end{equation}
	where $\vect\theta_\epsilon=\vect\theta_f\cup\vect\theta_\tau$, $f(y_{p}) \sim \mathcal{GP}(\mu_{f}(y_{p}),k _{f}(y_{p},y_{p}^{\prime});\vect{\theta}_{f})$ is a Gaussian process with hyperparameters $\vect{\theta}_{f}$ to model the mean $\mu_{f}(y_{p})$ and kernel $k _{f}(y_{p},y_{p}^{\prime})$, $\tau(y_{p}) \sim \mathcal{N}(0,\exp(g(y_{p})))$ is an input-dependent Gaussian noise with zero mean and variance $\exp(g(y_{p}))$, and $g(y_{p})\sim \mathcal{GP}(\mu_{g},k _{g}(y_{p},y_{p}^{\prime});\vect{\theta}_{g})$ is another Gaussian process with hyperparameters $\vect{\theta}_{g}$ to model the mean $\mu_{g}$ and kernel $k _{g}(y_{p},y_{p}^{\prime})$. %It is noted that a homoscedastic Gaussian process model can be readily derived if the Gaussian noise term in Eq.\eqref{hGPerrorcorrection} is independent of the input $y_{L}$, i.e., $\varepsilon(y_{L})=\varepsilon \sim \mathcal{N}(0,\sigma_{\varepsilon}^{2})$, where $\sigma_{\varepsilon}$ is a constant standard deviation. 
	
	The introduction of the heteroscedastic Gaussian noise is essential in this study because the mapping from $y_p$ to $y$ can be one-to-many. The use of heteroscedastic Gaussian process comes with a price of increasing the number of hyperparameters, and the analytical marginal likelihood and prediction equations for the homoscedastic Gaussian process are no longer useful \cite{lazaro2013retrieval,rogers2020probabilistic,kim2023estimation}. The hyperparameters $\vect{\theta}_{f}$ and $\vect{\theta}_{g}$ can be approximated by maximizing the following lower bound of the exact marginal likelihood \cite{lazaro2013retrieval}:
	\begin{equation}\label{likelihoodlowerbound}
		F(\vect{\mu} ,\vect{\Sigma} )=\textrm{log}f_{\mathcal{N}}(\vect{\varepsilon};\vect{0},\vect{K}_{f}+\vect{R})-\frac{1}{4}\textrm{tr}(\vect{\Sigma})-\textrm{KL}(f_{\mathcal{N}}(\vect{g};\vect{\mu},\vect{\Sigma})|| f_{\mathcal{N}}(\vect{g};\mu_{g}\vect{1},\vect{K}_{g})) \,,
	\end{equation}
	where $\vect{\mu}$ and $\vect{\Sigma}$ are the variational mean vector and covariance matrix to be determined alongside with the hyperparameters $\vect{\theta}_{f}$ and $\vect{\theta}_{g}$; $\vect{\varepsilon}=[{\varepsilon}^{(1)},{\varepsilon}^{(2)},...]\tr$ are from the training set $\mathcal{D}_\epsilon$; $\vect{0}$ and $\vect{1}$ are vectors of zeros and ones; $\vect{K}_{f}$ and $\vect{K}_{g}$ are respectively the covariance matrices of the Gaussian processes $f(y_{p})$ and $g(y_{p})$; {$\vect{R}$ is a diagonal matrix with diagonal entries $R_{ii}=\exp(\mu_{i}-\Sigma_{ii}/2)$}; $f_{\mathcal{N}}$ denotes the probability density function of a multivariate Gaussian distribution; $\textrm{tr}(\cdot)$ denotes matrix trace; and $\textrm{KL}(\cdot||\cdot)$ denotes the Kullback–Leibler divergence between two probability density functions. 
	
	Once the hyperparameters $\vect{\theta}_{f}$ and $\vect{\theta}_{g}$ are estimated, the predictions for the mean and variance of $\epsilon(y_{p}^{\ast})$ at $y_{p}^{\ast}$ are obtained from \cite{lazaro2013retrieval}:
	\begin{equation}\label{PredictionMean}
		\mu_{{\epsilon}}(y^{\ast}_{p})=\vect{k}_{f\ast}\tr(\vect{K}_{f}+\vect{R})^{-1}\vect{\varepsilon}\,,
	\end{equation}
	and
	\begin{equation}\label{PredictionSigma}
		\sigma_{{\epsilon}}^{2}(y_{p}^{\ast})=\exp\left(\vect{k}_{g\ast}\tr(\vect{\Lambda}-\tfrac{1}{2}\vect{I})\vect{1}+\mu_{g}+\frac{k_{g\ast\ast}-\vect{k}_{g\ast}\tr(\vect{K}_{g}+\vect{\Lambda}^{-1})^{-1}\vect{k}_{g\ast}}{2} \right)+k_{f\ast\ast}-\vect{k}_{f\ast}\tr(\vect{K}_{f}+\vect{R})^{-1}\vect{k}_{f\ast}\,,
	\end{equation}
	respectively, where $k_{f\ast\ast}=k_{f}(y_{p}^{\ast},y_{p}^{\ast})$, $k_{g\ast\ast}=k_{g}(y_{p}^{\ast},y_{p}^{\ast})$, $\vect{k}_{f\ast}=[k_{f}(y_{p}^{(1)},y_{p}^{\ast}),k_{f}(y_{p}^{(2)},y_{p}^{\ast}),...]\tr$ and \linebreak ${\vect{k}_{g\ast}=[k_{g}(y_{p}^{(1)},y_{p}^{\ast}),k_{g}(y_{p}^{(2)},y_{p}^{\ast}),...]{\tr}}$, $\vect{I}$ is the identity matrix, and $\vect{\Lambda}$ is a positive semi-definite diagonal matrix introduced to re-parameterize $\vect{\mu}$ and $\vect{\Sigma}$ in a reduced order.
	
\begin{figure}[H]
		\centering
		\includegraphics[scale=0.5]{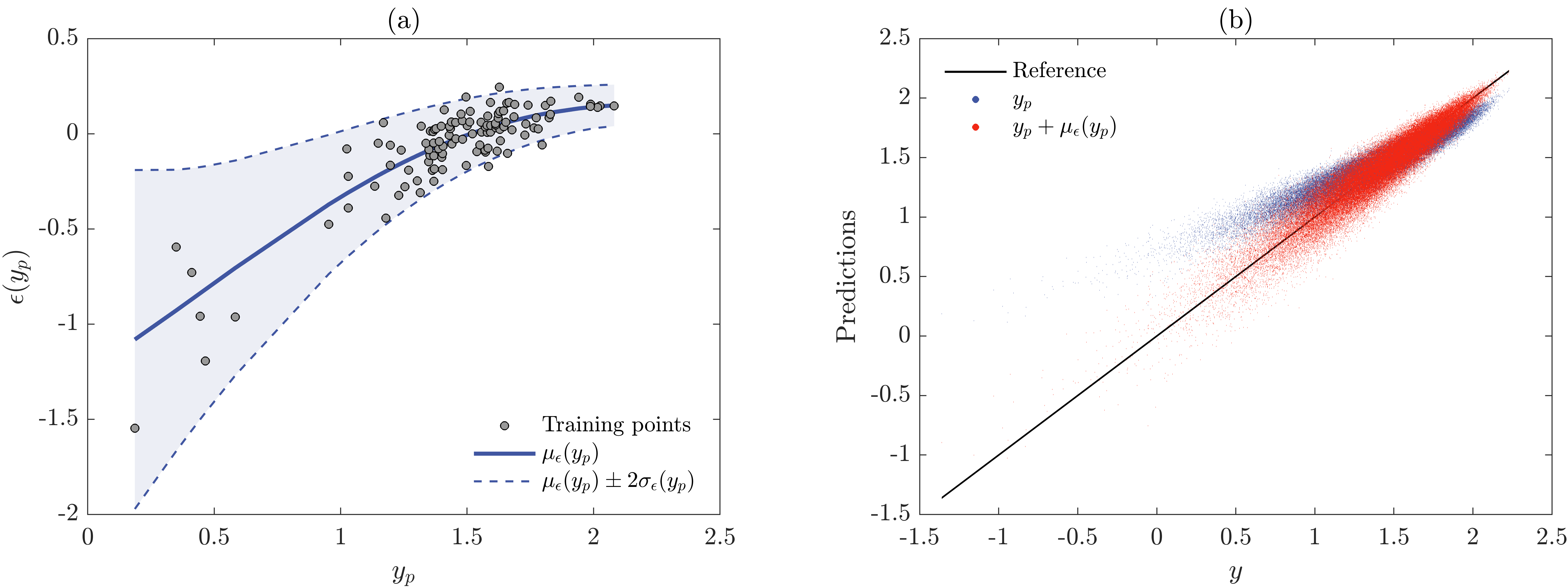}
		\caption{\textbf{Heteroscedastic Gaussian process for error correction.} \textit{Following Figure \ref{Fig:Figure1}, for $\rho=0.946$, a heteroscedastic Gaussian process is trained to fit the error. It is seen that the heteroscedastic errors are well-captured, and the heteroscedastic Gaussian process corrects the surrogate model predictions. {The mean absolute relative errors between the surrogate model predictions and the true responses, before and after error correction, are $12.19\%$ and $7.08\%$, respectively. }}}
		\label{Fig:Figure2}
	\end{figure}

	The mean $\mu_{{\epsilon}}(y^{\ast}_{p})$ can be used as the prediction of the surrogate modeling error at $y_{p}^{\ast}$, and $\sigma_{{\epsilon}}^{2}(y_{p}^{\ast})$ quantifies the uncertainty of this prediction. The {latter} quantity is desirable in active learning. It is worth noting that the roles of the heteroscedastic Gaussian process involve quantifying the noise and correcting the (mean) predictions of the physics-based surrogate model (see Figure \ref{Fig:Figure2}); it cannot reduce the noise. %or increase the degree of correlation between the high- and low-fidelity model responses.
	
	\subsection{Adaptive training by active learning}\label{Sec:ActiveLearning}
	\noindent Given an initial training set $\mathcal{D}=\{(\vect x^{(i)},\mathcal{M}(\vect x^{(i)}))\}_{i=1}^{N_0}\equiv\mathcal{D}_{\vect x}\times\mathcal{D}_y$, we sequentially train the physics-based surrogate model $y_p=(\mathcal{M}_p\circ\psi)(\vect x;\vect\theta_p)$ and the error correction function $\epsilon(y_p;\vect\theta_\epsilon)$ to obtain the initial surrogate model $\hat y=(\mathcal{M}_p\circ\psi)(\vect x;\vect\theta_p)+\epsilon(y_p;\vect\theta_\epsilon)$. Subsequently, we initiate an active learning process to iteratively enrich the training set $\mathcal{D}$ and update the surrogate model. Hereafter, we develop an active learning process tailored to the physics-data-driven surrogate modeling. 
 
 %The optimized low-fidelity physical model $\mathcal{M}_{L}(\vect{x};\vect{\theta}_{L}^{\ast})$ and the corresponding response samples $\mathcal{Y}_{L}=\{ y_{L,i}=\mathcal{M}_{L}(\vect{x}_{i};\vect{\theta}_{L}^{\ast}),i=1,2,...,n_0 \}$ can be obtained using the ingredients presented in Section \ref{Sec:optimization}. Thereafter, based on $\mathcal{Y}_{H}$ and $\mathcal{Y}_{L}$, the heteroscedastic Gaussian process model of error corrections $\hat{\epsilon}(y_{L})$, along with the prediction mean $\mu_{\hat{\epsilon}}(y_{L})$ and variance $\sigma_{\hat{\epsilon}}^{2}(y_{L})$, can be constructed using the ingredients presented in Section \ref{Sec:ErrorCorrection}. The above two steps can yield the coupled physics-data-driven surrogate model regarding the initial training points. However, such a coupled surrogate model is typically inadequate for rare event simulation associated with low-probability estimation, e.g., small failure probability estimation, because the initial training points may all fall into the high-probability region. In view of this, certain active learning techniques \cite{bichon2008efficient,echard2011ak} are desirable for adaptively identifying the training points around the vicinity of the limit state that considerably contributes to the failure probability. 
	
	%Let $\mathcal{X}=\{\vect{x}^{(i)}\}_{i=1}^N$, $N\gg N_0$, denote the candidate training set formed by sampling from $f_{\vect X}$. 
 
 Provided with the current surrogate model, we first define a critical learning region $\Omega_c$ as:
	\begin{equation}\label{LearningFun1}
		\Omega_c=\left \{ \vect{x}\in\rn:\frac{\left |y_{p}+\mu_{{\epsilon}}(y_{p})\right |}{\sigma_{{\epsilon}}(y_{p})} \leqslant \delta ,\,\,\, y_{p}=(\mathcal{M}_{p}\circ\psi)(\vect{x};\vect{\theta}_{p}) \right \}\,,
	\end{equation}
	where $\delta>0$ is a cutoff value for the learning region, and $\mu_\epsilon(y_p)$ and $\sigma_\epsilon(y_p)$ are respectively the mean and standard deviation of the Gaussian process error correction at $y_p$. 
 
 Next, we generate a candidate training set  ${\mathcal{X}}_c=\{\vect x^{(i)}\}_{i=1}^N$ through: 
 \begin{equation}\label{LearningKernel}
     \vect X^{(i)}\sim\1{\Omega_c}{\vect x}f_{\vect X}(\vect x)\,,
 \end{equation}
 where $\mathbb{I}_{\Omega_c}:\rn\mapsto\{0,1\}$ is an indicator function for $\{\vect x\in\Omega_c\}$ and the normalizing constant for the density $\1{\Omega_c}{\vect x}f_{\vect X}(\vect x)$ is omitted for simplicity. To reduce surrogate model simulations and improve the scalability toward low probability estimations, the sequential Monte Carlo \cite{papaioannou2016sequential,wang2019hamiltonian,xian2024relaxation} is used to generate ${\mathcal{X}}_c$. 
 
 Given ${\mathcal{X}}_c$, the  next training point $\vect{x}^{\ast}$ is identified by:
	\begin{equation}\label{LearningFun2}
		\vect{x}^{\ast}=\mathop{\arg\max}\limits_{\vect{x}\in {\mathcal{X}}_c} \left ( \mathop{\min}\limits_{y_p^\prime \in{\mathcal{Y}_p}}\left\|{y}_p-{y}_p^\prime \right \| \right ),\,\,\, y_{p}=(\mathcal{M}_{p}\circ\psi)(\vect{x};\vect{\theta}_{p})\,,
	\end{equation}
 where $\mathcal{Y}_p=\{y_{p}=(\mathcal{M}_{p}\circ\psi)(\vect{x};\vect{\theta}_{p}):\vect x\in\mathcal{D}_{\vect x}\}$ is a set of $y_p$ predictions from existing training samples. This equation picks the point in $\mathcal{X}_c$ that differs (in terms of $y_p$) the most from the existing training points, aiming to achieve sparsely distributed training points in the space of $y_p$. This learning criterion is a relaxation of the classic U-function-based learning \cite{echard2011ak} through introducing additional, seemingly redundant, distance-based selection to enforce sparsity. This relaxation is necessary because the prediction uncertainty term $\sigma_\epsilon(y_p)$ of the U-function, which influences the ``sparsity" of the U-function-based training point selection, is not only affected by the distance to existing training points, but also contributed, sometimes dominantly, by the inherent noise of the surrogate model. In the conventional application \cite{echard2011ak,echard2013combined,huang2016assessing} of active learning-based Gaussian process modeling, the training data does not have inherent stochastic noise, and the distance-based selection in Eq.~\eqref{LearningFun2} can be unnecessary. Incidentally, the constructions of Eq.~\eqref{LearningFun1} and Eq.~\eqref{LearningKernel} are still meaningful even for noise-free scenarios, because they facilitate the use of efficient sampling methods to identify the low-probability critical region to propose training points. 
	
Provided with the next training point $\vect{x}^{\ast}$, the original model is simulated to obtain $y^{\ast}=\mathcal{M}(\vect{x}^{\ast})$, the training set $\mathcal{D}$ is augmented to include $(\vect{x}^{\ast},y^{\ast})$, and the surrogate model is updated. {The learning process is stopped if the mean prediction is stable and/or the prediction uncertainty is small  \cite{dubourg2011reliability,moustapha2022active}. In our context, the inherent noise originated from the imperfection of the physics-based surrogate model cannot be reduced through training. Additionally, the magnitude of this inherent noise varies with the problem. Thus, the conventional prediction uncertainty-based stopping criterion is not suitable. To address this, we adopt a stopping criterion that monitors the stability of prediction uncertainty, expressed as follows:
	\begin{equation}\label{Convergence}
		\frac{\left | \hat{P}_{\Delta}^{(m+1)}-\hat{P}_{\Delta}^{(m)} \right |}{\hat{P}_{\Delta}^{(m+1)}}\leqslant \eta, \,\,\, m=0,1,...\,,
	\end{equation}
where $\hat{P}_{\Delta}^{(m)}=\hat{P}^{+}-\hat{P}^{-}$ measures the prediction uncertainty, and $\hat{P}^{+}$ and $\hat{P}^{-}$ are respectively the upper- and lower- confidence limits of the rare event probability at the $m$-th learning step, estimated from the surrogate models $y_{p}+\mu_{{\epsilon}}(y_{p})-\sigma_{{\epsilon}}(y_{p})$ and $y_{p}+\mu_{{\epsilon}}(y_{p})+\sigma_{{\epsilon}}(y_{p})$, respectively. The mechanism of the stopping criterion is that the epistemic uncertainty from insufficient training is minimized when $\hat{P}_{\Delta}^{(m)}$ is stabilized; the remaining component of $\hat{P}_{\Delta}^{(m)}$ is the aleatory uncertainty that cannot be reduced by training, unless another physics-based surrogate model is used. }
	%\begin{equation}\label{Convergence}
	%	\eta=\frac{\left | \hat{P}_{f}^{(m+1)}-\hat{P}_{f}^{(m)} \right |}{\hat{P}_{f}^{(m+1)}}\leqslant \eta_c \,,
%	\end{equation}
	%where $\hat{P}_{f}^{(m)}$ and $\hat{P}_{f}^{(m+1)}$ are the failure probability estimations obtained by the Monte Carlo simulation of the coupled surrogate models at the $m$-th and ($m+$1)-th learning steps, respectively, and $\eta_c$ is the specified tolerance of convergence.
	
	The implementation details for the training of the coupled physics-data-driven surrogate model are presented in Appendix \ref{Fig:Appendix1}. Due to the presence of inherent noises, there is no guarantee that the probability prediction from the surrogate model will converge to the true value. To address this issue, a final importance sampling step will be introduced in the following section.
	
	\section{Importance sampling using the coupled physics-data-driven surrogate model}\label{Sec:ImportanceSampling}
	\noindent The target rare event probability can be formulated as
	\begin{equation}\label{FailureProbability}
		P=\int _{\vect{x}\in \rn}\1{\leq0}{\mathcal{M}(\vect{x})}f_{\vect{X}}(\vect{x})\mathrm{d}\vect{x}\,.
	\end{equation}
	%where $\vect{X}$ is assumed to be a random vector of $D$ independent standard Gaussian variables, $f_{\vect{X}}(\vect{x})$ is the joint probability density function of $\vect{X}$, and $\mathbb{I}(\mathcal{M}_{H}(\vect{x})\geqslant  b)$ is the binary indicator function that gives 1 if $\mathcal{M}_{H}(\vect{x})\geqslant  b$ and 0 otherwise.
 The importance sampling rewrites the integral in Eq.~\eqref{FailureProbability} by introducing an importance density $h(\vect{x})$, i.e.,
	\begin{equation}\label{ImportanceSampling}
		P=\int _{\vect{x}\in \rn}\1{\leq0}{\mathcal{M}(\vect{x})}\frac{f_{\vect{X}}(\vect{x})}{h(\vect{x})}h(\vect{x})\mathrm{d}\vect{x}\,.
	\end{equation}
	The support of the importance density $h(\vect{x})$ should cover the rare event, and the optimal importance density is \cite{rubinstein1998modern}
	\begin{equation}\label{ImportanceDensityHF}
		h^{\ast }(\vect{x})=\frac{\1{\leq0}{\mathcal{M}(\vect{x})}f_{\vect{X}}(\vect{x})}{P}\,.
	\end{equation}
	Due to the high correlation between the original and surrogate model responses, it is tempting to use the following importance density from the surrogate model as an approximation for the optimal density. 
	\begin{equation}\label{ImportanceDensityLF}
		\hat{h}(\vect{x})=\frac{\1{\leq0}{\hat{\mathcal{M}}(\vect x;\vect\theta_p,\vect\theta_{\epsilon})}f_{\vect{X}}(\vect{x})}{\hat{P}}\,,
	\end{equation}
	where $\hat{\mathcal{M}}(\vect x;\vect\theta_p,\vect\theta_{\epsilon})\equiv(\mathcal{M}_{p}\circ\psi)(\vect{x};\vect{\theta}_{p})+\mu_{{\epsilon}}((\mathcal{M}_{p}\circ\psi)(\vect{x};\vect{\theta}_{p});\vect\theta_\epsilon)$ is the integrated surrogate model, and $\hat{P}$ is its rare event probability. However, there is no guarantee that the target rare event is an improper subset of the support of $\hat h(\vect{x})$. Therefore, $\hat h(\vect{x})$ cannot be directly employed as an importance density. One remedy is to replace the binary indicator function in Eq.~\eqref{ImportanceDensityLF} by a smooth approximation \cite{papaioannou2016sequential,wang2024optimized}, so that $\hat h(\vect{x})$ becomes nonzero everywhere. This approach requires the tuning of a relaxation parameter associated with the smooth approximation of the indicator function. 
    
\begin{figure}[H]
		\centering
		\includegraphics[scale=0.5]{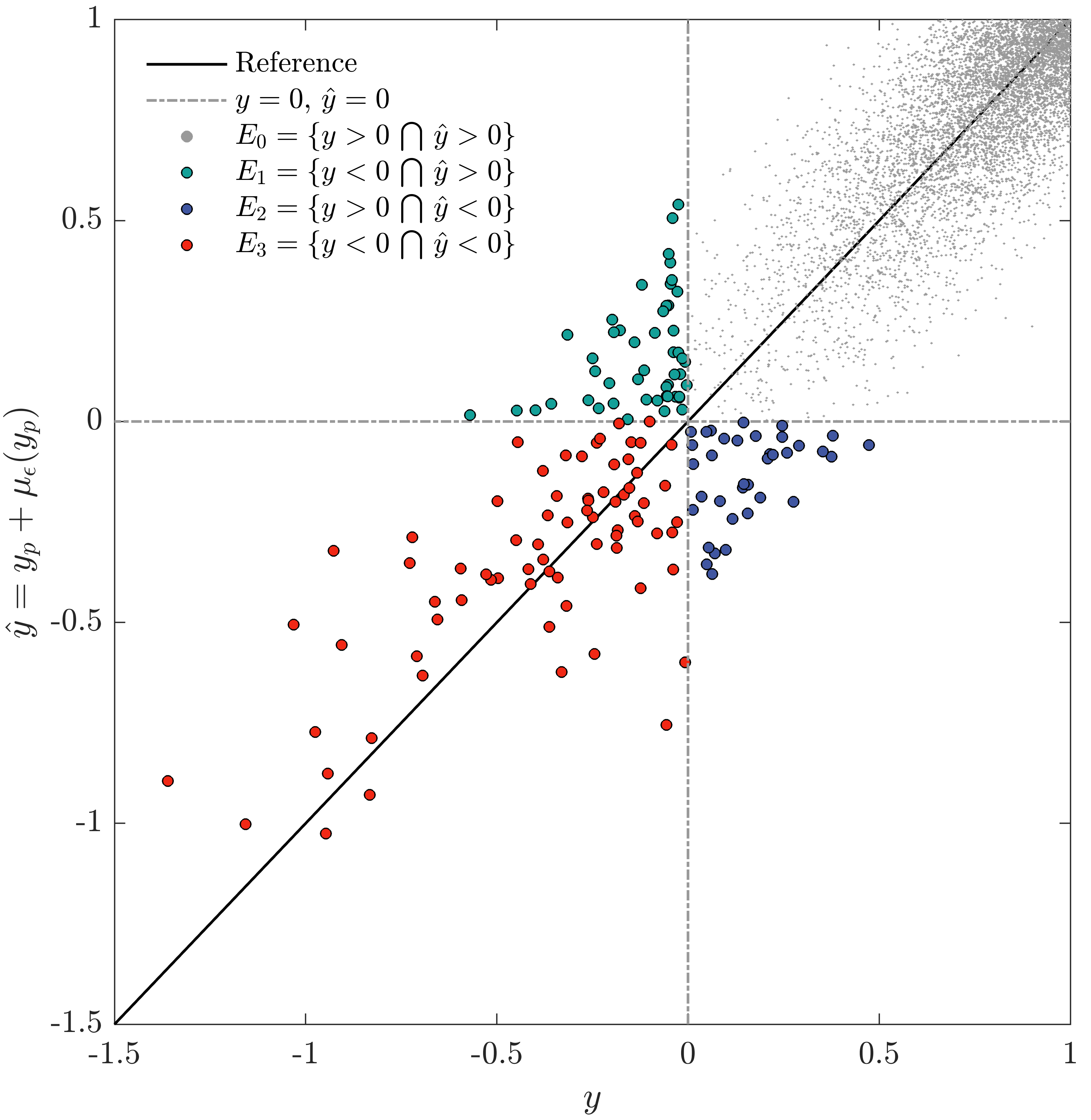}
		\caption{\textbf{Importance sampling for the coupled physics-data-driven surrogate model.} \textit{The true rare event from the original model is $E_1\cup E_3$, and the ``surrogate" rare event from the surrogate model is $E_2\cup E_3$. Eq.~\eqref{Identity} is derived from the identity $\Prob{E_1\cup E_3}=\Prob{E_2\cup E_3}\frac{\Prob{E_3}/{\Prob{E_2\cup E_3}}}{\Prob{E_3}/{\Prob{E_1\cup E_3}}}$. It follows that if $\Prob{E_1}=\Prob{E_2}$, $P=\hat P$, i.e., $\hat P$ can be accurate even when the overlap $E_3$ is not significant. This condition can be met if $\hat Y$ is an unbiased estimator of $Y$ for the region $E_1\cup E_2\cup E_3$, supporting the proposed approach of training a Gaussian process to correct the bias in the critical region identified by active learning}.}
		\label{Fig:Figure3}
	\end{figure}

    An alternative remedy, which is adopted in this paper, is to use the following identity derived from properties of conditional probability \cite{wang2024optimized}:
	\begin{equation}\label{Identity}
		P=c_P\cdot\hat{P}\equiv\frac{\int _{\vect{x}\in\rn}\1{\leq0}{\mathcal{M}(\vect{x})}\hat h(\vect{x})\mathrm{d}\vect{x}}{\int _{\vect{x}\in \rn}\1{\leq0}{\hat{\mathcal{M}}(\vect x;\vect\theta_p,\vect\theta_{\epsilon})}h^{\ast }(\vect{x})\mathrm{d}\vect{x}}\,\hat{P}\,,
	\end{equation}
	where $c_P$ is the correction factor for the surrogate model-based rare event probability estimation. The numerator represents the conditional probability {of} the original rare event given the surrogate rare event, while the denominator represents the conditional probability {of} the surrogate rare event given the original rare event. Ideally, the correction factor $c_P$ should be close to $1$; therefore, $c_P$ can be used as a performance measure of the surrogate model. The conditional probabilities in Eq.~\eqref{Identity} can be estimated by Monte Carlo simulation with Markov Chain Monte Carlo algorithms \cite{neal2011mcmc,papaioannou2015mcmc} to generate samples from $h^*(\vect x)$ and $\hat h(\vect x)$. For a well-trained surrogate model, the overlap between $h^*(\vect x)$ and $\hat h(\vect x)$ is expected to be significant; consequently, the computational cost to estimate $c_P$ would be small. However, theoretically speaking, Eq.~\eqref{Identity} implies that the surrogate model solution $\hat P$ can be highly accurate even when the surrogate rare event does not largely overlap with the actual rare event; this is illustrated by Figure \ref{Fig:Figure3}. Finally, the implementation details of the importance sampling using Eq.~\eqref{Identity} are presented in Appendix \ref{Fig:Appendix2}. 
		
	\section{Numerical examples}\label{Sec:Application}
	\noindent In this section, we will investigate (1) a static problem of a linear elastic cantilever beam with material properties modeled by a Gaussian random field, (2) a dynamic problem of a nonlinear viscous damper under white noise excitation, and (3) a dynamic problem of a multi-degree-of-freedom hysteretic system under white noise excitation. Three schemes are investigated to construct physics-based surrogate models. Specifically, homogenization of material properties is considered in the first example, statistical linearization is used in the second example, and relaxation of the numerical solver is adopted in the third example.
	
	\subsection{Example 1: A linear elastic cantilever beam }\label{Sec:Applicationone}
\begin{figure}[H]
		\centering
		\includegraphics[scale=0.055]{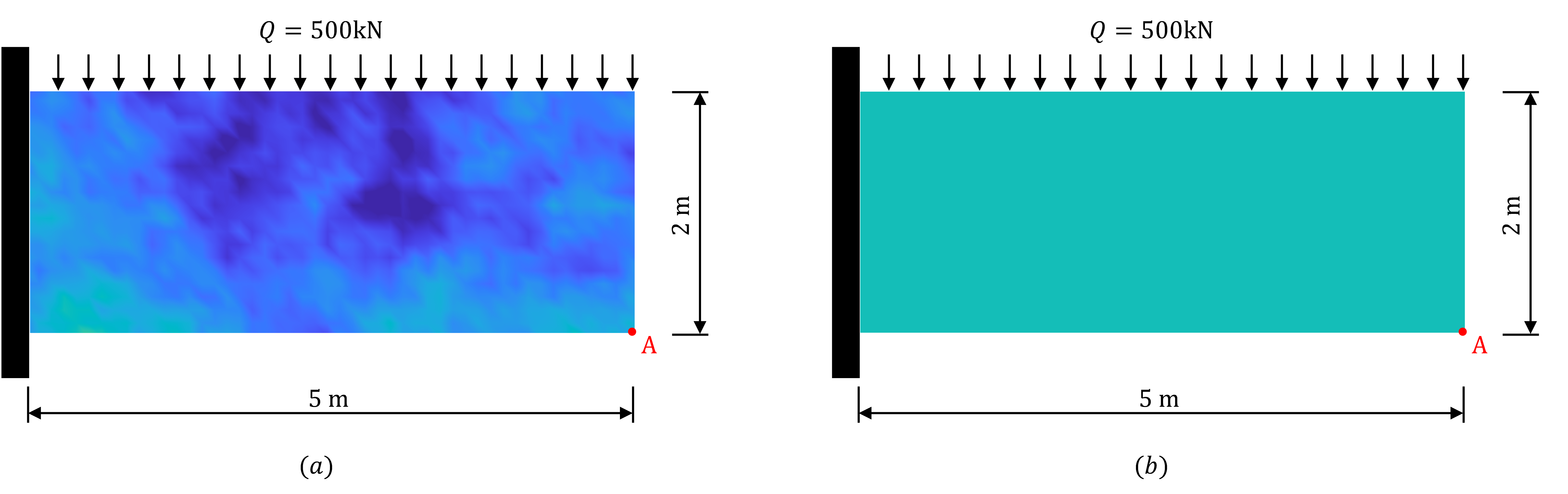}
		\caption{\textbf{{Physics-based surrogate model of a linear elastic cantilever beam.}} \textit{(a) original model: the spatial domain of the beam is discretized into $50\times 20=1000$ four-node quadrilateral elements. The elastic modulus of the beam is characterized by a homogeneous random field $E(x,y)$ with a discretization of $1000$ Gaussian random variables compatible with the spatial discretization; (b) physics-based surrogate model: the whole beam is represented by a single four-node quadrilateral element. The elastic modulus is assumed to be a homogenization of the random field, i.e., $E_{p}=a_{E}\bar{E}+b_{E}$, where $a_{E}$ and $b_{E}$ are tunable parameters, and $\bar{E}$ is the average of the $1000$ Gaussian random variables.}}
		\label{Fig:Example1_fig1}
	\end{figure}

	\noindent Consider a linear elastic cantilever beam of length $5$m and width $2$m subjected to 20 evenly spaced concentrated loads on the upper edge with magnitudes $Q=500\mathrm{kN}$, shown in Figure \ref{Fig:Example1_fig1}. The Poisson ratio of the beam is $\nu=0.3$. The reference/original computational model is represented by a discretization of the beam into $50\times20=1000$ four-node quadrilateral elements, and the elastic modulus of the beam is characterized by a homogeneous random field $E(x,y)$ with a discretization of $1000$ Gaussian random variables compatible with the spatial discretization. The autocorrelation function of the Gaussian random field is modeled by the isotropic exponential model $R_{E}(\Delta _{x},\Delta _{y})=\exp\left (-\sqrt{\Delta _{x}^2+\Delta _{y}^2}/l  \right )$ with the correlation length $l=10\mathrm{m}$. The mean value and standard deviation of the marginal Gaussian distribution are $\mu _{E}=200\mathrm{GPa}$ and $\sigma _{E}=30\mathrm{GPa}$, respectively. The physics-based surrogate model is constructed via a simple homogenization of the random field of the elastic modulus: 
 \begin{equation}
  E_{p}=a_{E}\bar{E}+b_{E}\,,   
 \end{equation}
 where $\{a_{E},b_{E}\}=\vect\theta_p$ are tunable parameters of the physics-based surrogate model, and $\bar{E}$ represents the average of the $1000$ Gaussian random variables. The initial values of $a_{E}$ and $b_{E}$ are set to be $1$ and $0$, respectively. The rare event of interest is defined as the vertical displacement of point A exceeds a prescribed threshold of $0.0032\mathrm{m}$, expressed by
 \begin{equation}
   \{y=0.0032-d_A\leq0\}\,,  
 \end{equation}
where $d_A$ is the vertical displacement of point A.  

\begin{figure}[H]
		\centering
		\includegraphics[scale=0.4]{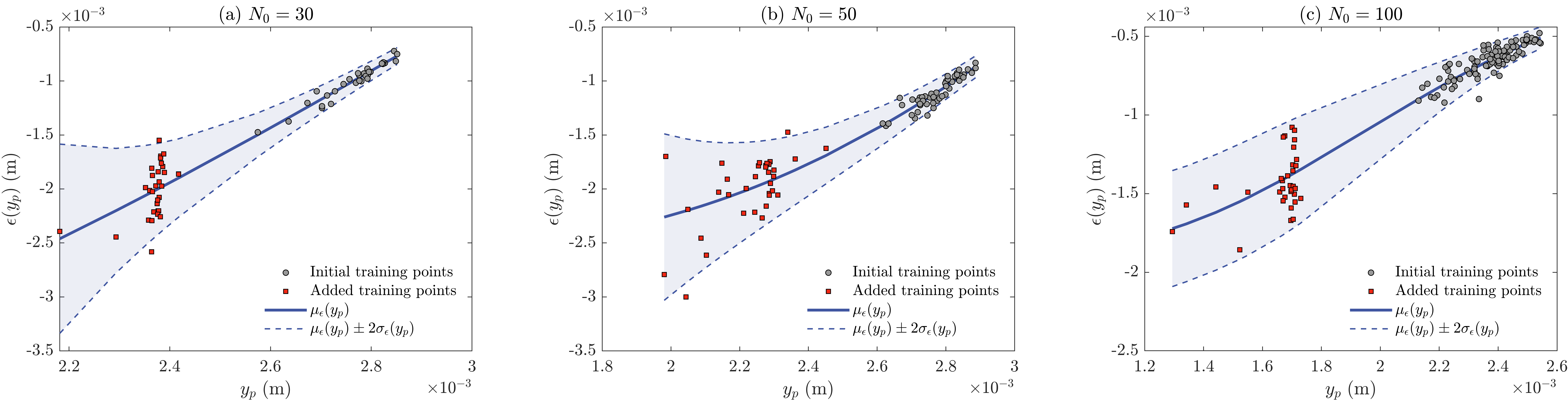}
		\caption{\textbf{{Heteroscedastic Gaussian process model of the error correction at the final learning stage for Example 1.}} \textit{{From left to right, the Gaussian process models are obtained from $30(\text{initial})+31(\text{active learning})$, $50(\text{initial})+34(\text{active learning})$, and $100(\text{initial})+33(\text{active learning})$ training data, respectively.} It is seen that the heteroscedastic Gaussian process model can capture the noisy errors.}}
		\label{Fig:Example1_fig2}
	\end{figure}

\begin{figure}[H]
		\centering
		\includegraphics[scale=0.4]{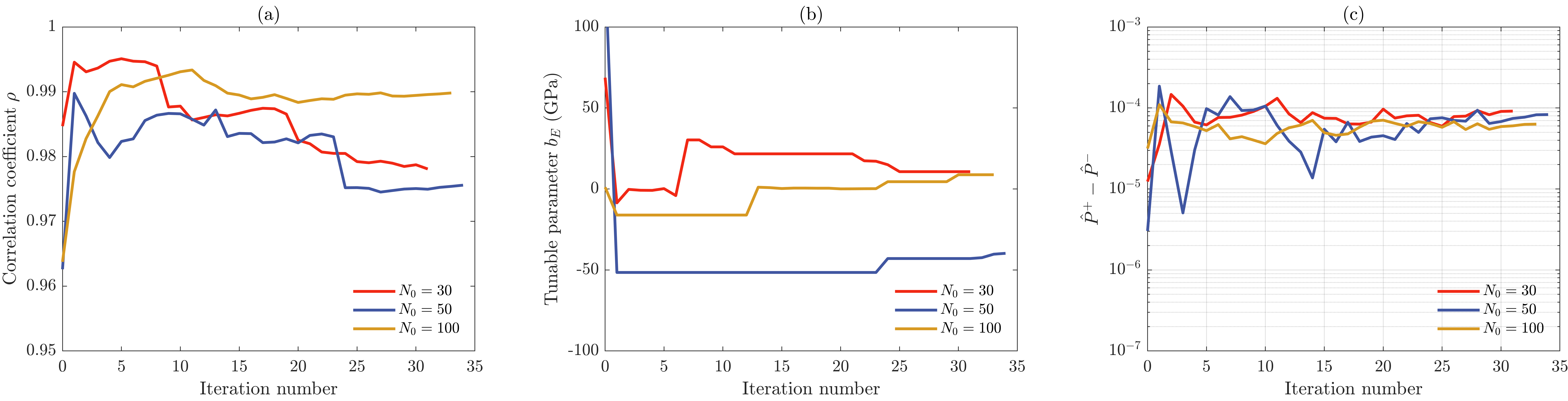}
		\caption{\textbf{{Optimization histories of the proposed method for Example 1.}} \textit{{(a) the correlation coefficient ${\rho}$ between $Y_p$ and $Y$ at the training points achieves $0.9781$, $0.9756$, and $0.9898$ for $N_0=30$, $N_0=50$, and $N_0=100$, respectively; (b) the tunable parameter $b_{E}$ for the physics-based surrogate model converges to $10.7\mathrm{GPa}$, $-39.7\mathrm{GPa}$, and $8.8\mathrm{GPa}$ for $N_0=30$, $N_0=50$, and $N_0=100$, respectively; (c) the prediction uncertainties stabilize after $30$ active learning steps.}}}%the probability estimation $\hat{P}$ converges to $3.3020\times10^{-5}$, $3.3225\times10^{-5}$, and $2.4380\times10^{-5}$ for $N_0=30$, $N_0=50$, and $N_0=100$, respectively.} The reference solution from subset simulation with $4.6\times10^{4}$ samples is $3.6130\times10^{-5}$.}}
		\label{Fig:Example1_fig3}
	\end{figure}

\begin{figure}[H]
		\centering
		\includegraphics[scale=0.4]{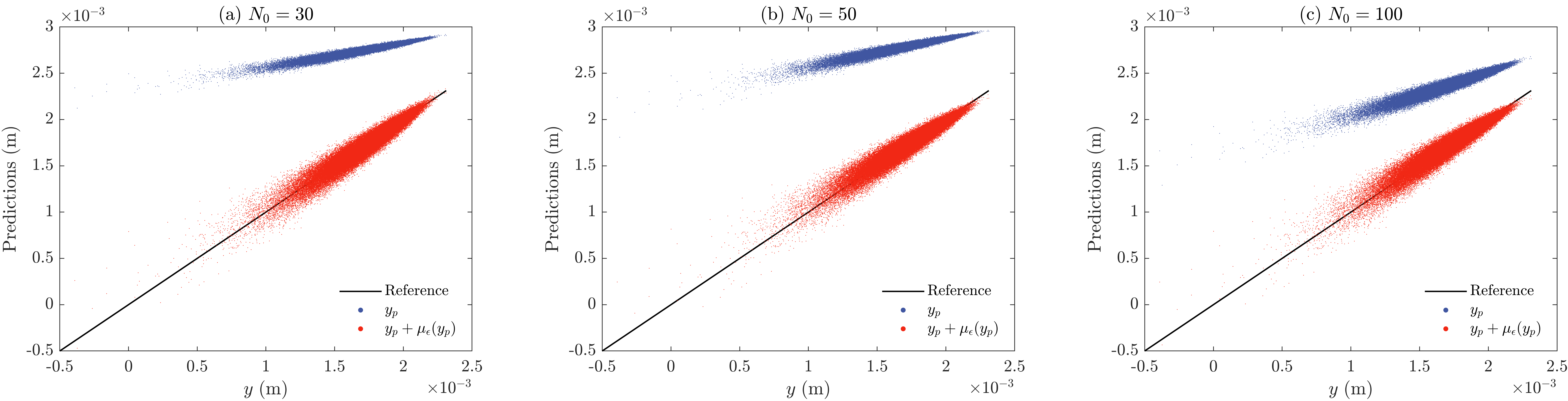}
		\caption{\textbf{{Response predictions of the physics-based surrogate model and the coupled physics-data-driven surrogate model for Example 1.}} \textit{The response prediction of the physics-based surrogate model $Y_p$ is highly correlated with the true responses $Y$, and the data-driven error correction improves the bias. {The mean absolute relative errors between the surrogate model predictions and the true responses, before and after error correction, are $65.12\%$ and $3.83\%$ for $N_0=30$, $67.39\%$ and $3.75\%$ for $N_0=50$, and $41.85\%$ and $3.74\%$ for $N_0=100$, respectively.}}}
		\label{Fig:Example1_fig4}
	\end{figure}

\begin{figure}[H]
		\centering
		\includegraphics[scale=0.5]{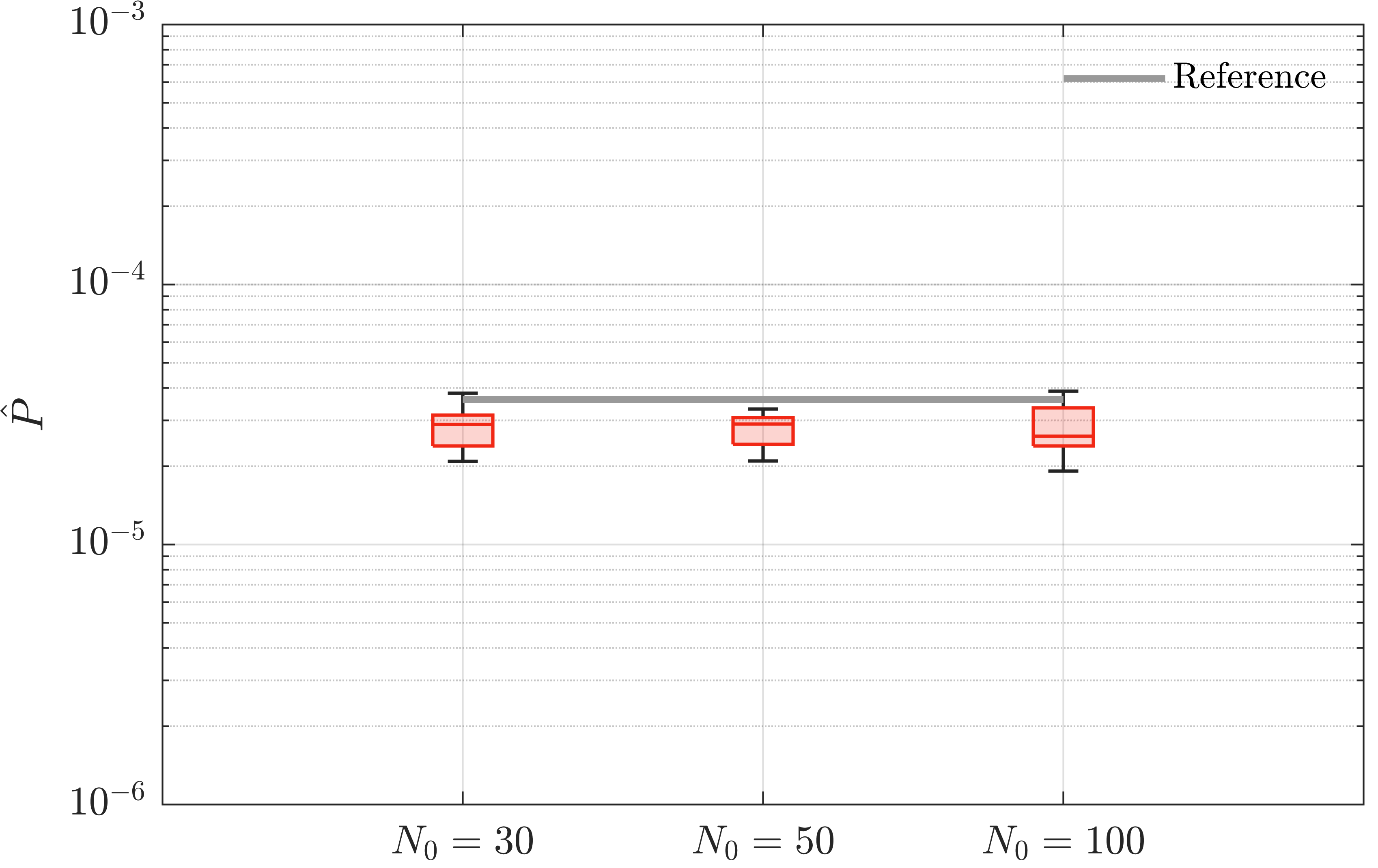}
		\caption{\textbf{{Probability estimations using different sizes of the initial training set for Example 1.}} \textit{{Each box plot is obtained using $10$ independent runs of the proposed surrogate modeling method.}}}
		\label{Fig:Example1_fig5}
	\end{figure}

{The performance of the proposed method is investigated across different sizes of the initial training set, $N_0=30,\,50,\,100$. We found that $30$ initial points are typically required for stable training. During the active learning process, around $33$ new training points are identified, insensitive to $N_0$. This is expected because the initial training set may not offer sufficient information on the rare event--starting from $30$ or $100$ training points may not make much difference.} The heteroscedastic Gaussian process model for the error correction at the final learning stage is shown in Figure \ref{Fig:Example1_fig2}. It is seen that the heteroscedastic Gaussian process can model the noisy errors. {The optimization histories of the correlation coefficient $\rho$, the parameter $b_{E}$, and the prediction uncertainty $\hat{P}_{\Delta}=\hat P^+-\hat P^-$ are shown in Figure \ref{Fig:Example1_fig3}. The scatter plots of $y_p$ and $y_p+\epsilon(y_p)$ against $y$ are shown in Figure \ref{Fig:Example1_fig4}}. It is seen that the response predictions of the physics-based surrogate model are highly correlated with the responses of the original model, and the data-driven error correction can improve the bias of the surrogate predictions. {To estimate the statistical variability of the proposed surrogate modeling method, the box plot of probability estimations using $10$ independent runs is shown in Figure \ref{Fig:Example1_fig5}. It is observed that the method has a small but noticeable bias for rare event probability estimations. Furthermore, increasing the number of initial training points may not improve the prediction bias and variability. Therefore, using $30$ initial training points seems to be ideal for the current problem. The first quartile, median, and third quartile of the rare event probability estimations are presented in Table \ref{tab:1}.} Finally, importance sampling is implemented to improve the probability estimation. To achieve a coefficient of variation of $5\%$, $400$ runs of the original model is typically required. The final probability estimate ranges from $3.5 \times 10^{-5}$ to $4.0 \times 10^{-5}$, with a correction factor $c_P$ between $1.1$ and $1.6$.

%{$3.7024\times10^{-5}$($3.4890\times10^{-5}$)($3.9980\times10^{-5}$)}. This indicates that the correction factor $c_P$ (recall Eq.~\eqref{Identity}) for the surrogate model solution is {$1.12$($1.05$)($1.64$)}. 

	\subsection{Example 2: An oscillator with nonlinear viscous damper}\label{Sec:Applicationtwo}
	\noindent Consider an oscillator with nonlinear viscous damper under a Gaussian white noise excitation with the equation of motion expressed as
	\begin{equation}\label{viscousoscillator}
		m\ddot{u}(t)+c\dot{u}(t)+ku(t)+c_{d}\mathrm{sign}(\dot{u}(t))\left | \dot{u}(t) \right |^{\alpha_{d}}=-m\ddot{u}_g(t)\,,
	\end{equation}
	where $m=3000\mathrm{kg}$, $c=3000\mathrm{N/(m/s)}$, and $k=3\times10^5\mathrm{N/m}$ are the mass, damping, and stiffness of the oscillator, respectively, $u(t)$, $\dot{u}(t)$, and $\ddot{u}(t)$ are the displacement, velocity, and acceleration of the oscillator, respectively, $c_{d}=800\mathrm{N/(m/s)}^{\alpha_{d}}$ and $\alpha_{d}=0.3$ are the damping coefficient and velocity exponent of the nonlinear viscous damper, respectively, $\mathrm{sign}(\cdot )$ is the sign function, and $\ddot{u}_g(t)$ is the acceleration excitation.
	
	The excitation has a duration of $15$ seconds. Using the spectral representation method \cite{shinozuka1991simulation}, the white noise can be discretized into a finite set of Gaussian variables:  
	\begin{equation}\label{Seismicexcitationrandomprocess}
		\ddot{u}_g(\vect{X},t)=\sum_{i=1}^{D/2}\sqrt{2S_{0}\Delta \omega }\left ( X_i\textrm{cos}(\omega_it)+ \bar{X}_i\mathrm{sin}(\omega_it)\right )\,,
	\end{equation}
	where $X_i$ and $\bar{X}_i$, $i=1,2,...,D/2$, are mutually independent standard normal variables, $D=1000$, $\Delta \omega=2\omega_{\max}/D$ is the frequency increment with $\omega_{\max}=25\pi$ being the upper cutoff
	angular frequency, $\omega_i=(i-0.5)\Delta\omega$ is the discretized frequency points, and $S_0=5\times10^{-3} \mathrm{m^2/s^3}$ is the intensity of the white noise.
	
	To obtain the physics-based surrogate model, the nonlinear equation of motion shown in Eq.~\eqref{viscousoscillator} is linearized into
	\begin{equation}\label{viscousoscillatorlinearized}
		m\ddot{u}(t)+(c+c_{e})\dot{u}(t)+ku(t)=-m\ddot{u}_g(t)\,,
	\end{equation}
	where $\{c_{e}\}=\vect\theta_p$ is the equivalent damping coefficient used as the tuning parameter of the physics-based surrogate model. The rare event of interest is defined as the peak absolute displacement of the oscillator exceeding a threshold of $0.06$m, expressed by
 \begin{equation}
  \left\lbrace y=0.06-\sup_{t\in[0,15]}|u(t)|\leq0\right\rbrace\,. 
 \end{equation}
 The initial value of $c_{e}$ for optimization is determined by the traditional statistical linearization method \cite{xian2020stochastic}. 
 
\begin{figure}[H]
		\centering
		\includegraphics[scale=0.4]{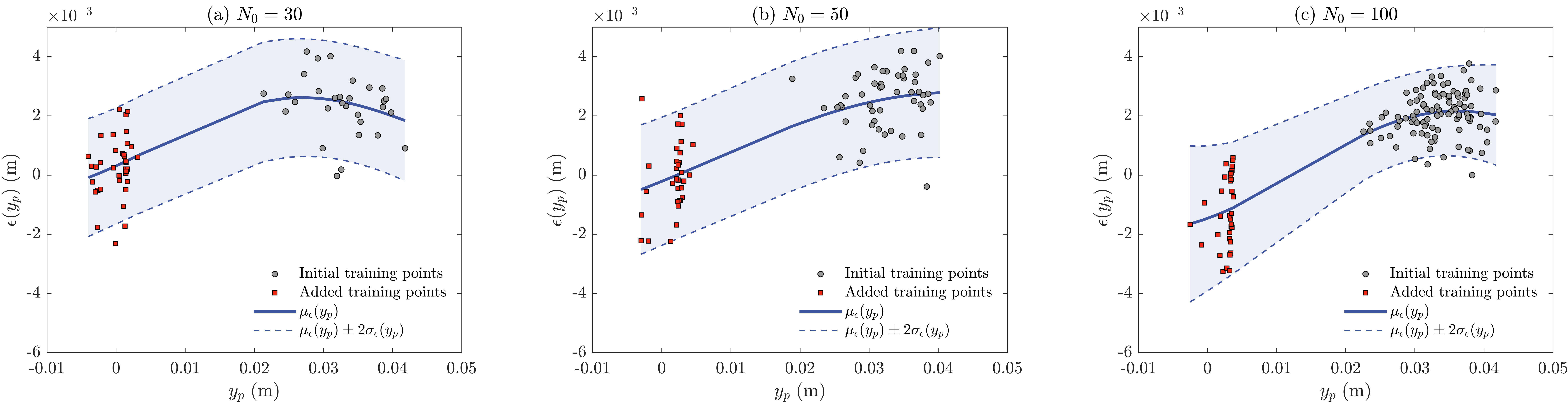}
		\caption{\textbf{{Heteroscedastic Gaussian process model of the error correction at the final learning stage for Example 2.}} \textit{{From left to right, the Gaussian process models  are obtained using $30(\text{initial})+37(\text{active learning})$, $50(\text{initial})+32(\text{active learning})$, and $100(\text{initial})+34(\text{active learning})$ training data, respectively.} The heteroscedastic Gaussian process model captures the noisy errors.}}
		\label{Fig:Example2_fig1}
	\end{figure}

\begin{figure}[H]
		\centering
		\includegraphics[scale=0.4]{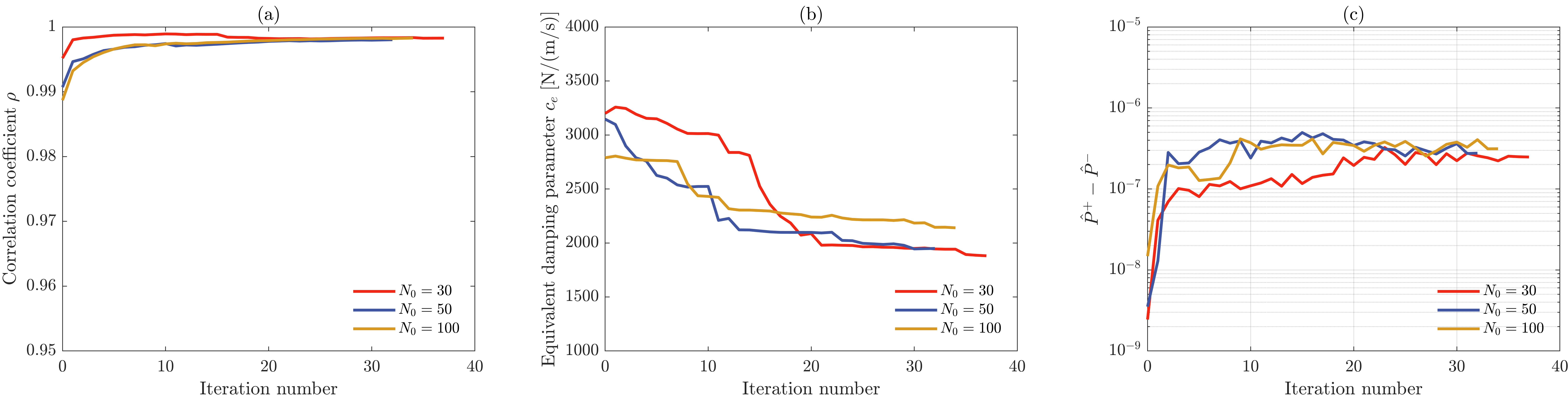}
		\caption{\textbf{{Optimization histories of the proposed method for Example 2.}} \textit{{(a) the correlation coefficient ${\rho}$ between $Y_p$ and $Y$ at the training points achieves $0.9983$, $0.9980$, and $0.9983$ for $N_0=30$, $N_0=50$, and $N_0=100$, respectively; (b) the equivalent damping parameter $c_e$ for the physics-based surrogate model stops at $1881.1\mathrm{N/(m/s)}$, $1949.0\mathrm{N/(m/s)}$, and $2140.2\mathrm{N/(m/s)}$ for $N_0=30$, $N_0=50$, and $N_0=100$, respectively; (c) the prediction uncertainties stabilize after 30 active learning steps }.}}%the probability estimation $\hat{P}$ converges to $2.0598\times10^{-7}$, $2.2905\times10^{-7}$, and $2.2575\times10^{-7}$ for $N_0=30$, $N_0=50$, and $N_0=100$, respectively.} The reference solution of subset simulation with $6.49\times10^{4}$ samples is $2.9540\times10^{-7}$
		\label{Fig:Example2_fig2}
	\end{figure}

\begin{figure}[H]
		\centering
		\includegraphics[scale=0.4]{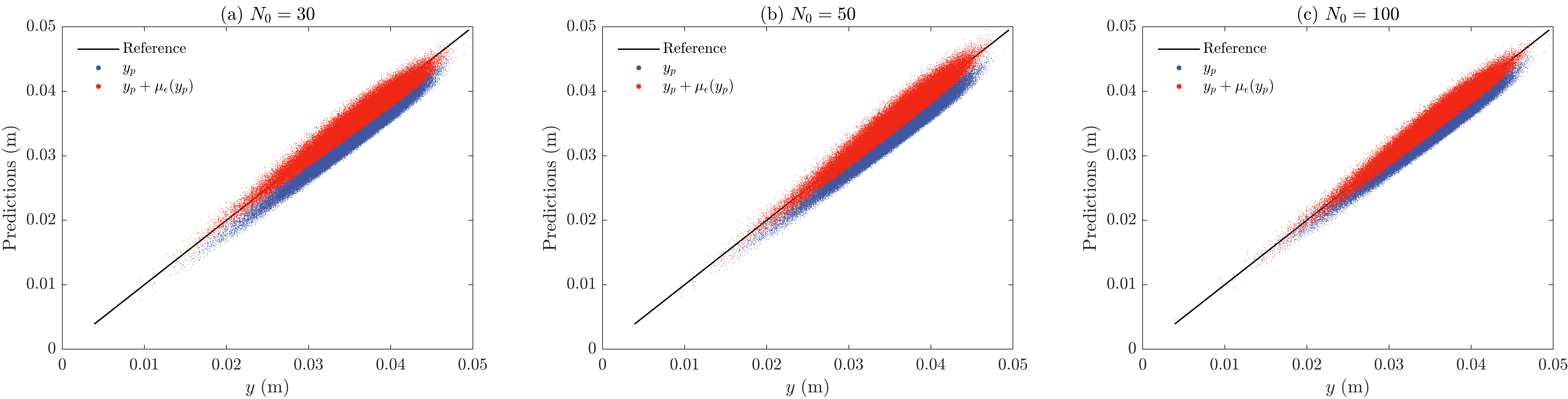}
		\caption{\textbf{{Response predictions of physics-based surrogate model and coupled physics-data-driven surrogate model for Example 2.}} \textit{The response prediction of the physics-based surrogate model $Y_p$ is highly correlated with the true responses $Y$, and the data-driven error correction improves the bias. {The mean absolute relative errors between the surrogate model predictions and the true responses, before and after error correction, are $7.48\%$ and $2.33\%$ for $N_0=30$, $7.03\%$ and $2.23\%$ for $N_0=50$, and $5.80\%$ and $1.99\%$ for $N_0=100$, respectively.}}}
		\label{Fig:Example2_fig3}
	\end{figure}

\begin{figure}[H]
		\centering
		\includegraphics[scale=0.5]{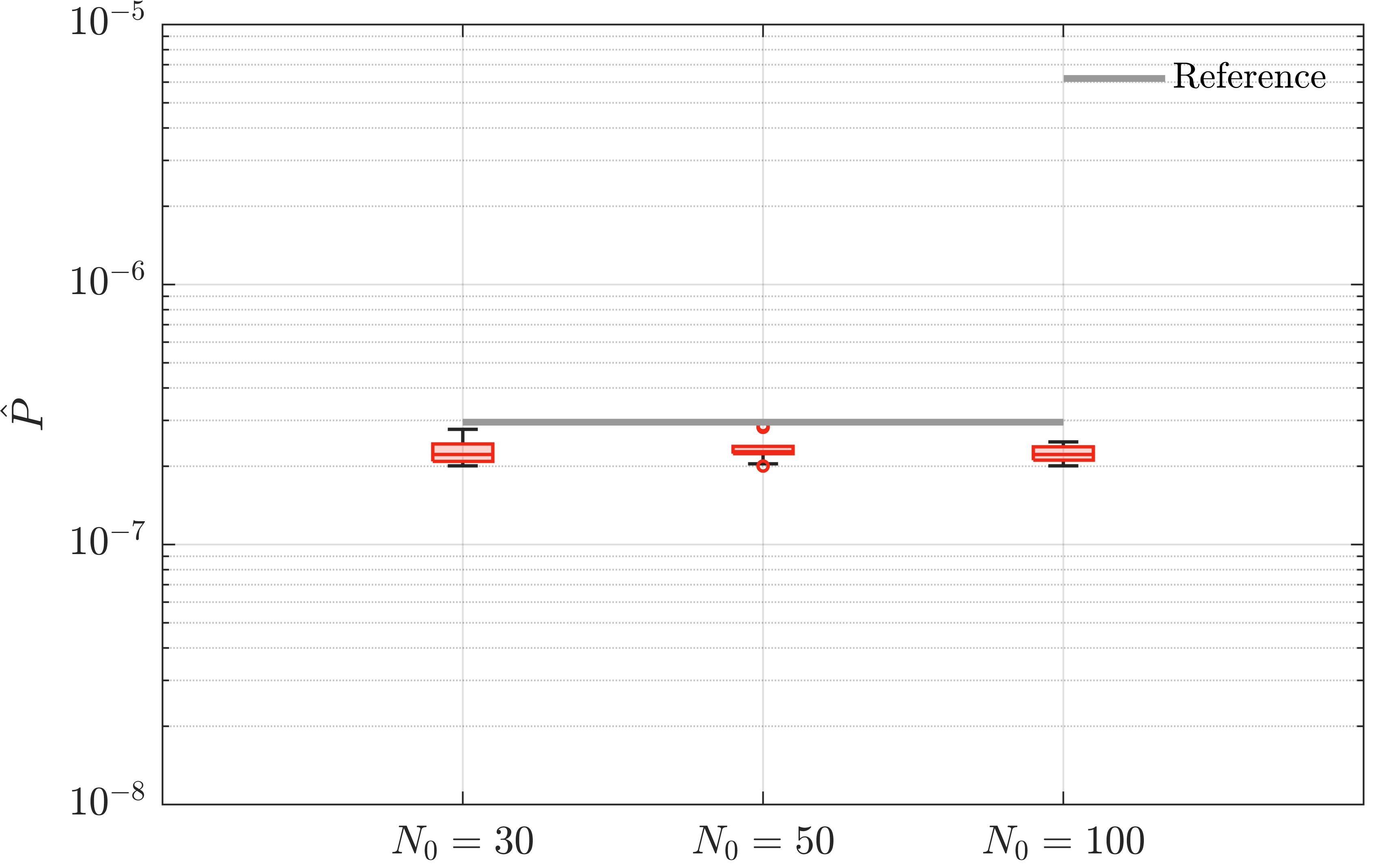}
		\caption{\textbf{{Probability estimations using different sizes of the initial training set for Example 2. }} \textit{{Each box plot is obtained using $10$ independent runs of the proposed surrogate modeling method.}}}
		\label{Fig:Example2_fig4}
	\end{figure}

 The coupled physics-data-driven surrogate model is trained using {$30,\,50,\,\text{and }100$ initial samples and  around $34$ active learning samples}. The heteroscedastic Gaussian process model for error correction at the final learning stage is shown in Figure \ref{Fig:Example2_fig1}. The optimization histories of the correlation coefficient ${\rho}$, the equivalent damping coefficient $c_{e}$, and the prediction uncertainty  {$\hat P_{\Delta}=\hat{P}^{+}-\hat{P}^{-}$} are shown in Figure \ref{Fig:Example2_fig2}.  The scatter plots of $y_p$ and $y_p+\epsilon(y_p)$ against $y$ are shown in Figure \ref{Fig:Example2_fig3}. {The box plot of probability estimations using 10 independent runs of the proposed surrogate modeling method is shown in Figure \ref{Fig:Example2_fig4}, and the first quartile, median, and third quartile of the probability estimations are presented in Table \ref{tab:1}. Applying importance sampling, the final probability estimate ranges from $2.6 \times 10^{-7}$ to $2.8 \times 10^{-7}$, with a correction factor between $1.2$ and $1.4$, based on  $400$ runs of the original model.}
 
 %For the $N_0=30$ The correction factor $c_P$ in the importance sampling step is estimated to be {$1.35$($1.19$)($1.15$)} using an additional {$600$($400$)($400$)} runs of the original model, leading to a final probability estimate of {$2.7810\times10^{-7}$($2.7260\times10^{-7}$)($2.5960\times10^{-7}$)}.

 %The probability predicted by the coupled surrogate model is {$2.0598\times10^{-7}$($2.2905\times10^{-7}$)($2.2575\times10^{-7}$)}, which is close to the reference solution $2.9540\times10^{-7}$ obtained from subset simulation using $6.49\times 10^{4}$ samples. 
	
	\subsection{Example 3: A multi-degree-of-freedom  hysteretic system}\label{Sec:Applicationthree}
\begin{figure}[H]
		\centering
		\includegraphics[scale=0.03]{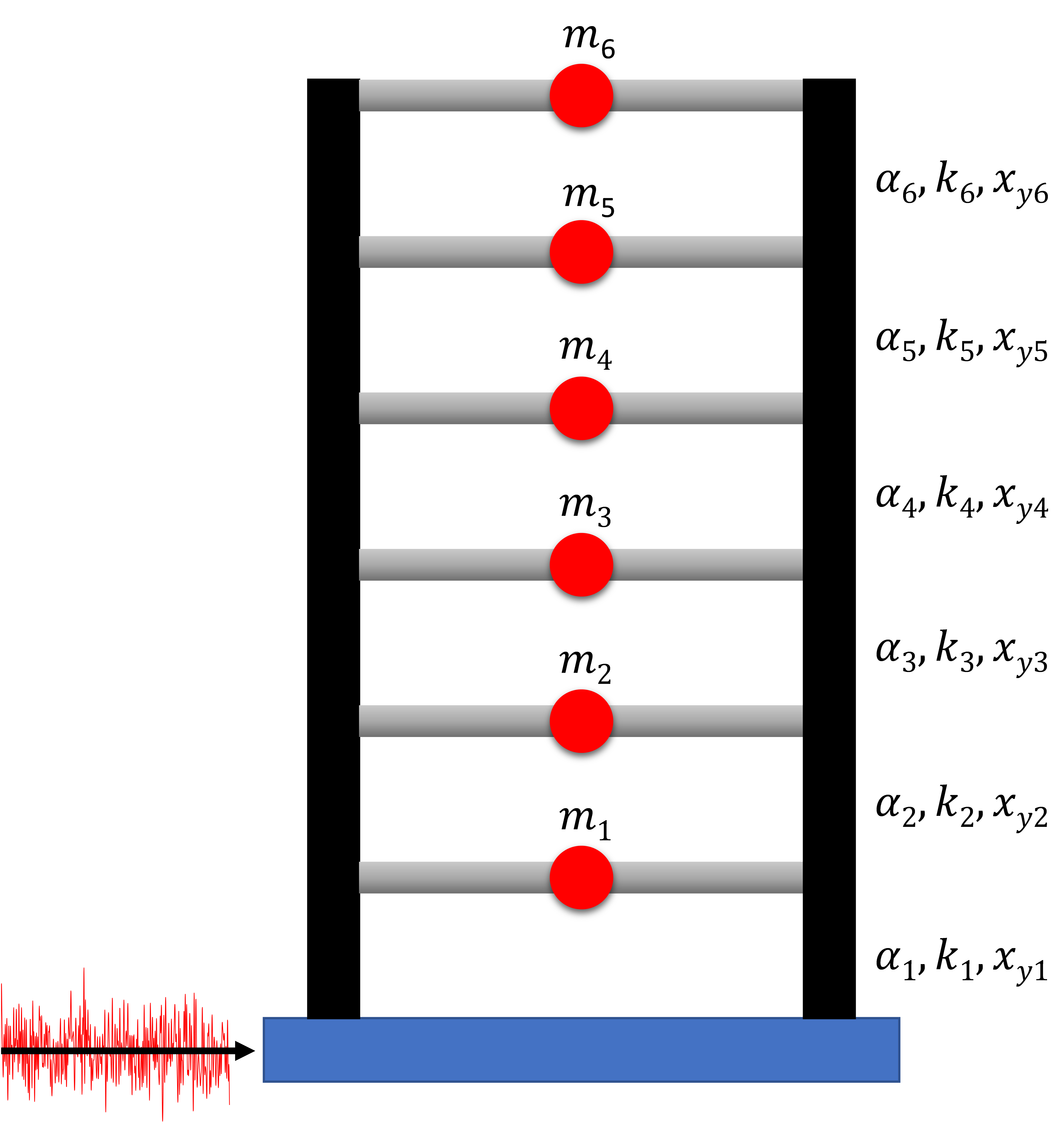}
		\caption{\textbf{A 6-degree-of-freedom shear-type  hysteretic system under ground motion excitation.} \textit{}}
		\label{Fig:Example3_fig1}
	\end{figure}

	\noindent Consider a 6-degree-of-freedom shear-type  hysteretic system under ground motion excitation, shown in Figure \ref{Fig:Example3_fig1}. The mass and initial stiffness of each storey are $m_{i}=8000\mathrm{kg}$ and $k_{i}=1\times10^7\mathrm{N/m}$, $i=1,2,...,6$, respectively. Rayleigh damping is assumed for the hysteretic system with a damping ratio of $5\%$ for the 1st and 6th mode of the system. The restoring force of each storey is described by the Bouc-Wen model \cite{wen1980equivalent} as follows:	
	\begin{equation}\label{BoucWen1}
		f_{i}(t)=\alpha_{i} k_{i}u_{i}(t)+(1-\alpha_{i})k_{i}z_{i}(t)\,,
	\end{equation}
	\begin{equation}\label{BoucWen2}
		\dot{z_{i}}(t)=g_{i}(\dot{u_{i}}(t),z_{i}(t))=\phi_{i} \dot{u_{i}}(t)-\varphi_{i} \left |\dot{u_{i}}(t)  \right |z_{i}(t)\left |z_{i}(t)  \right |^{\gamma_{i} -1}-\psi_{i}  \dot{u_{i}}(t)\left |z_{i}(t)  \right |^{\gamma_{i} }\,,
	\end{equation}
	where $\alpha_{i}=0.1$ is the stiffness reduction ratio of the $i$-th storey; $u_{i}(t)$, $\dot{u_{i}}(t)$, and $z_{i}(t)$ are the relative displacement, relative velocity, and the hysteretic displacement of the $i$-th storey, respectively; $\phi_{i}=1$, $\varphi_{i}=\psi_{i}=1/(2x_{yi}^{\gamma_{i}})$, and $\gamma_{i}=1$ are the shape parameters of the hysteresis loop; and $x_{yi}=1.25$mm is the yield displacement of the $i$-th storey. The excitation is assumed to be the same as that in Example 2, but the intensity of the white noise is set to $S_0=8.5\times10^{-4} \mathrm{m^2/s^3}$.
	
 The rare event of interest is defined as the peak absolute deformation among the six storeys exceeding a threshold of $0.015$m, expressed by
\begin{equation}
 \left\lbrace y=0.015-\max_{i\in\{1,2,...,6\}}\left(\sup_{t\in[0,15]}|u_i(t)|\right)\leq0\right\rbrace\,.   
\end{equation} 
We define the original and physics-based surrogate models based on solvers of the equation of motion shown in Eq.~\eqref{BoucWen2}. For the original model, Eq.~\eqref{BoucWen2} is solved by the implicit Euler algorithm:
	\begin{equation}\label{ImplicitEuler}
		z_{i}(t+\Delta t)=z_{i}(t)+g_{i}(\dot{u_{i}}(t+\Delta t),z_{i}(t+\Delta t))\Delta t\,,
	\end{equation}
	and for the surrogate model, Eq.~\eqref{BoucWen2} is solved by the explicit Euler algorithm:
	\begin{equation}\label{ExplicitEuler}
		z_{i}(t+\Delta t)=z_{i}(t)+g_{i}(\dot{u_{i}}(t),z_{i}(t))\Delta t\,,
	\end{equation}
	where $\Delta t$ is the time step. The explicit Euler algorithm is highly efficient but less accurate, and thus is ideal in constructing the physics-based surrogate model. The stiffness reduction ratio $\alpha_{i}$, initial stiffness $k_{i}$, and yield displacement $x_{yi}$ are set as tunable parameters for the physics-based surrogate model, i.e., $\vect\theta_p=\{\alpha_{i},k_{i},x_{yi}\}^6_{i=1}$. The initial values of the those parameters are set to be the same as the original model. 
	
\begin{figure}[H]
		\centering
		\includegraphics[scale=0.4]{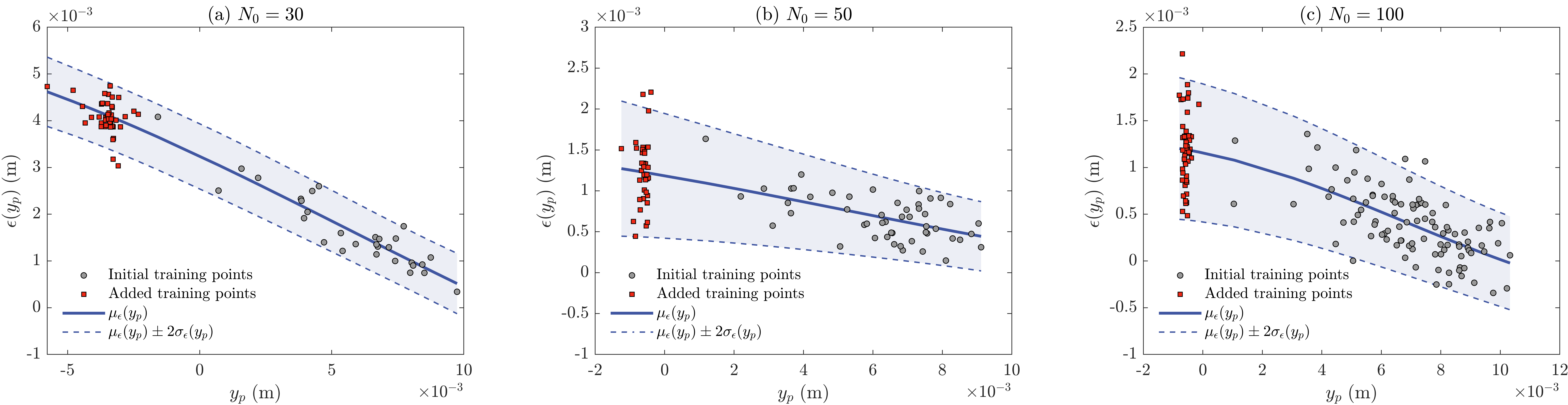}
		\caption{\textbf{{Heteroscedastic Gaussian process model of the error correction at the final learning stage for Example 3.}} \textit{{From left to right, the Gaussian process models are obtained using $30(\text{initial})+43(\text{active learning})$, $50(\text{initial})+37(\text{active learning})$, and $100(\text{initial})+52(\text{active learning})$ training data, respectively.} The heteroscedastic Gaussian process model captures the noisy errors}.}
		\label{Fig:Example3_fig2}
	\end{figure}

\begin{figure}[H]
		\centering
		\includegraphics[scale=0.4]{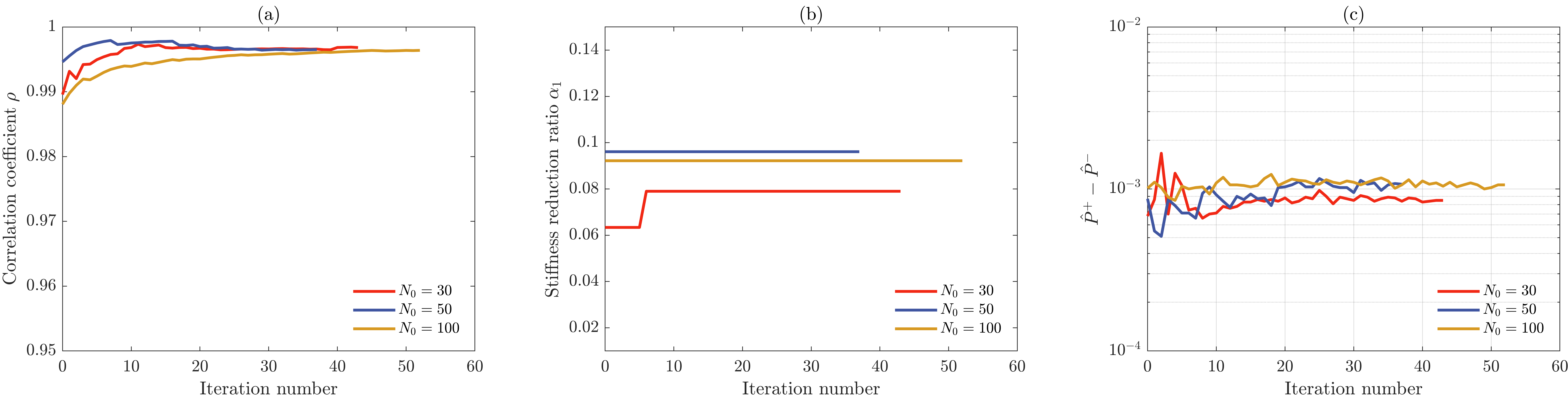}
		\caption{\textbf{{Optimization histories of the proposed method for Example 3.}} \textit{{(a) the correlation coefficient ${\rho}$ between $Y_p$ and $Y$ at the training points achieves $0.9968$, $0.9965$, and $0.9964$ for $N_0=30$, $N_0=50$, and $N_0=100$, respectively; (b) the first-storey stiffness reduction ratio $\alpha_{1}$ for the physics-based surrogate model converges to $0.0790$, $0.0961$, and $0.0922$ for $N_0=30$, $N_0=50$, and $N_0=100$, respectively; (c) the prediction uncertainties stabilize after $40$ active learning steps.}}}
		\label{Fig:Example3_fig3}
	\end{figure}

\begin{figure}[H]
		\centering
		\includegraphics[scale=0.4]{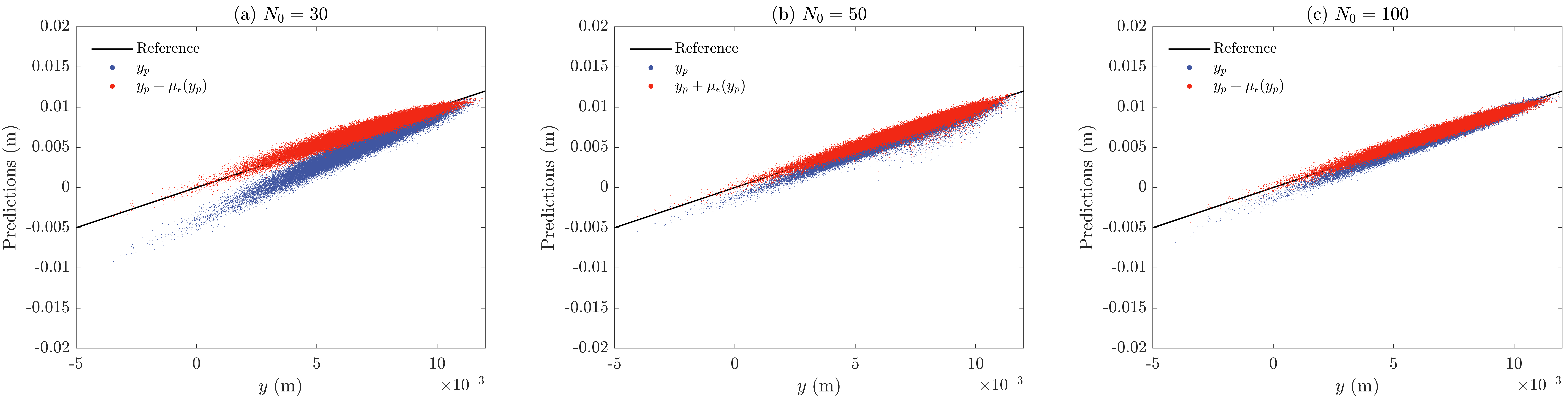}
		\caption{\textbf{{Response predictions of physics-based surrogate model and coupled physics-data-driven surrogate model for Example 3.}} \textit{The response prediction of the physics-based surrogate model $Y_p$ is highly correlated with the true responses $Y$, and the data-driven error correction corrects the bias. {The mean absolute relative errors between the surrogate model predictions and the true responses, before and after error correction,  are $188.29\%$ and $16.18\%$ for $N_0=30$, $62.64\%$ and $10.08\%$ for $N_0=50$, and $67.48\%$ and $19.09\%$ for $N_0=100$, respectively.}}}
		\label{Fig:Example3_fig4}
	\end{figure}

\begin{figure}[H]
		\centering
		\includegraphics[scale=0.5]{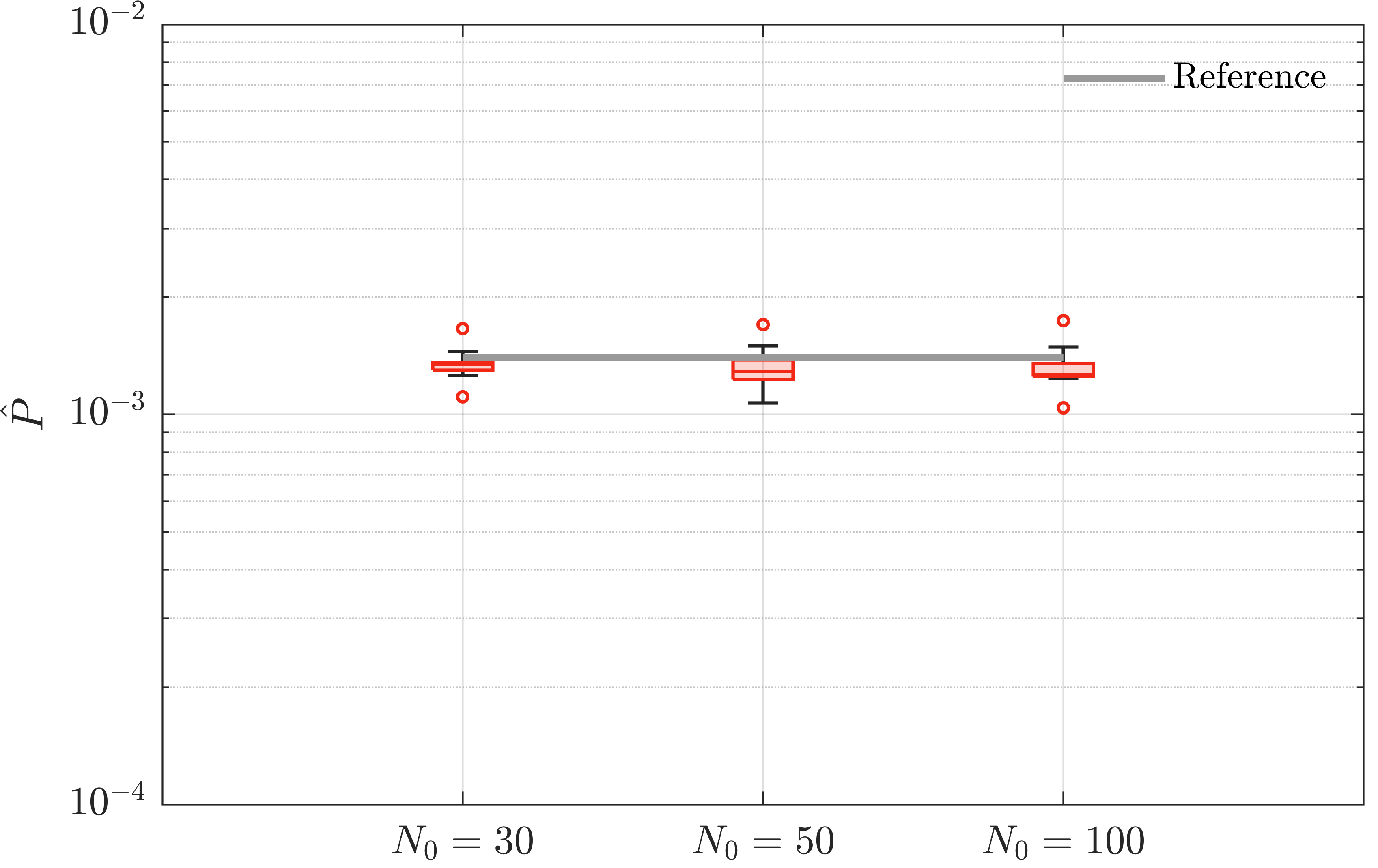}
		\caption{\textbf{{Probability estimations using different sizes of the initial training set for Example 3. }} \textit{{Each box plot is obtained using $10$ independent runs of the proposed surrogate modeling method.}}}
		\label{Fig:Example3_fig5}
	\end{figure}

	The coupled physics-data-driven surrogate model is trained using {$30$, $50$, and $100$ initial samples and around $45$ active learning samples}. The heteroscedastic Gaussian process model for error correction at the final learning stage is shown in Figure \ref{Fig:Example3_fig2}. The optimization histories of the correlation coefficient ${\rho}$, the first-storey stiffness reduction ratio $\alpha_{1}$, and the prediction uncertainty {$\hat P_{\Delta}=\hat{P}^{+}-\hat{P}^{-}$} are shown in Figure \ref{Fig:Example3_fig3}. Figure \ref{Fig:Example3_fig4} confirms  the accuracy of the coupled surrogate model. {Figure \ref{Fig:Example3_fig5} shows the box plot of probability estimations using $10$ independent runs of the proposed surrogate modeling method. Table \ref{tab:1} reports the first quartile, median, and third quartile of the probability estimations.} %It is worth noting that, in this example, the tunable parameters undergo milder change in the iteration histories because the physics-based surrogate model is characterized by $18$ parameters compared with only $1$ or $2$ parameters in the previous examples.

 %{where the mean probability prediction is $1.37\times10^{-3}$ and the reference solution from MCS using $10^{5}$ samples is $1.40\times10^{-3}$}. 
	
%\begin{table}[H]
%			\centering
%			\caption{{Probability estimations from the proposed surrogate modeling method (10 independent runs).}}
%			\begin{tabular}{c c c c c c}
	%			\toprule
   %              {}  & {} & \multicolumn{3}{c}{Surrogate model solution} & Reference\\
	%			Case  & $N_0$ & First quartile & Median & Third quartile & \\ \midrule
				
				%
	%			 {} & 30 & $2.3930\times 10^{-5}$ & $2.8900\times 10^{-5}$ & $3.1460\times 10^{-5}$ \\
				%
	%			Example 1 & 50 & $2.4320\times 10^{-5}$ & $2.9065\times 10^{-5}$ & $3.0790\times 10^{-5}$ & $3.6130\times 10^{-5}$ \\
                %
%				{} & 100 & $2.3940\times 10^{-5}$ & $2.6080\times 10^{-5}$ & $3.3520\times 10^{-5}$ \\ \midrule
				%
	%			{} & 30 & $2.0900\times 10^{-7}$ & $2.2175\times 10^{-7}$ & $2.4402\times 10^{-7}$ \\
				%
%				Example 2 & 50 & $2.2330\times 10^{-7}$ & $2.2729\times 10^{-7}$ & $2.3844\times 10^{-7}$ & $2.9540\times 10^{-7}$ \\
    	      %
%				{} & 100 & $2.1102\times 10^{-7}$ & $2.2153\times 10^{-7}$ & $2.3770\times 10^{-7}$ \\ \midrule
				%
%				{} & 30 & $1.3\times10^{-3}$ & $1.3\times10^{-3}$ & $1.4\times10^{-3}$ \\
				%
	%			Example 3 & 50 & $1.2\times10^{-3}$ & $1.3\times10^{-3}$ & $1.4\times10^{-3}$ & $1.4\times10^{-3}$ \\
   %             {} & 100 & $1.3\times10^{-3}$ & $1.3\times10^{-3}$ & $1.4\times10^{-3}$ \\
	%			\bottomrule
%			\end{tabular}
%			\label{tab:1}
%		\end{table}

  \begin{table}[H]
			\centering
			\caption{{Probability estimations from the proposed surrogate modeling method (10 independent runs).}}
			\begin{tabular}{c c c c c c}
				\toprule
                 {}  & {} & \multicolumn{3}{c}{Surrogate model solution} & \\
				Case  & $N_0$ & First quartile & Median & Third quartile & Reference \\ \midrule

				 {} & 30 & $2.4\times 10^{-5}$ & $2.9\times 10^{-5}$ & $3.1\times 10^{-5}$ \\
				Example 1 & 50 & $2.4\times 10^{-5}$ & $2.9\times 10^{-5}$ & $3.1\times 10^{-5}$ & $3.6\times 10^{-5}$ \\
				{} & 100 & $2.4\times 10^{-5}$ & $2.6\times 10^{-5}$ & $3.4\times 10^{-5}$ \\ \midrule
				{} & 30 & $2.1\times 10^{-7}$ & $2.2\times 10^{-7}$ & $2.4\times 10^{-7}$ \\
				Example 2 & 50 & $2.2\times 10^{-7}$ & $2.3\times 10^{-7}$ & $2.4\times 10^{-7}$ & $3.0\times 10^{-7}$ \\
				{} & 100 & $2.1\times 10^{-7}$ & $2.2\times 10^{-7}$ & $2.4\times 10^{-7}$ \\ \midrule
				{} & 30 & $1.3\times10^{-3}$ & $1.3\times10^{-3}$ & $1.4\times10^{-3}$ \\
				Example 3 & 50 & $1.2\times10^{-3}$ & $1.3\times10^{-3}$ & $1.4\times10^{-3}$ & $1.4\times10^{-3}$ \\
                {} & 100 & $1.3\times10^{-3}$ & $1.3\times10^{-3}$ & $1.4\times10^{-3}$ \\
				\bottomrule
			\end{tabular}
			\label{tab:1}
		\end{table}

        {
	\section{Additional remarks and future directions}\label{Sec:lim}
		\subsection{Multiple critical domains of the rare event}
            \noindent The data-driven error correction may not perform well in the presence of multiple critical domains with different response behaviors. Here, the crux is not the ``multiple critical domains" per se, but the ``different response behaviors" within these domains. For instance, the first-passage probability problems examined in Example 2 and 3 are series systems with numerous modes corresponding to the times at which crossing events occur. However, when the response process reaches stationarity, the rare crossing events become homogeneous across different time points. As a result, the error correction function does not need to discern the specific time point of the crossing, since these events have similar statistical properties. In this context, a one-dimensional error correction function is effective. In contrast, if the computational model exhibits different behaviors across multiple critical domains for the rare event of interest, a one-dimensional error correction approach may introduce noise due to its inability to distinguish the details of the critical modes. A potential solution is to incorporate additional output variables to differentiate between critical domains and train distinct error functions for each domain.
            
            \subsection{Selection of physics-based surrogate models}
            \noindent Large noisy errors can be induced by the aforementioned mechanism, but another source is the inherent limitations of the selected physics-based surrogate model. We can reduce the inherent noise by using a more detailed surrogate model, but a trade-off between accuracy and efficiency is inevitable. Further studies are needed to standardize the selection of physics-based surrogate models for various classes of problems.
       
       \subsection{Alternative importance sampling strategies}
            \noindent The importance sampling Eq.~\eqref{Identity} can be reformulated into the following expression:
            \begin{equation}\label{Identity1}
		P=c_P\cdot\hat{P}\equiv\frac{\int _{\vect{x}\in\rn}\1{\leq0}{\mathcal{M}(\vect{x})}h_{\mathcal{M}\cup\hat{\mathcal{M}}}(\vect{x})\mathrm{d}\vect{x}}{\int _{\vect{x}\in \rn}\1{\leq0}{\hat{\mathcal{M}}(\vect x;\vect\theta_p,\vect\theta_{\epsilon})}h_{\mathcal{M}\cup\hat{\mathcal{M}}}(\vect{x})\mathrm{d}\vect{x}}\,\hat{P}\,,
	\end{equation}
where the importance density $h_{\mathcal{M}\cup\hat{\mathcal{M}}}$ is associated with the union of the original and surrogate rare events, expressed by:
            \begin{equation}
		h_{\mathcal{M}\cup\hat{\mathcal{M}}}(\vect x)\defi\frac{\1{\leq0}{\min(\mathcal{M}(\vect x),\hat{\mathcal{M}}(\vect x;\vect\theta_p,\vect\theta_{\epsilon}))}}{\Prob{\min(\mathcal{M}(\vect X),\hat{\mathcal{M}}(\vect X;\vect\theta_p,\vect\theta_{\epsilon}))\leq0}}f_{\vect X}(\vect x)\,.
	\end{equation}
Compared to Eq.~\eqref{Identity}, this alternative formulation allows for the use of a single set of samples from $h_{\mathcal{M}\cup\hat{\mathcal{M}}}$ to estimate both the numerator and the denominator. On the other hand, since the union event is larger than individual ones, running Markov Chain Monte Carlo on the union increases the likelihood of biased estimation for the correction factor $c_P$. From our preliminary tests, we have not observed a decisive performance gain from using this alternative formulation. Further investigations and tests are needed.

            %The noise level of data-driven error correction is vital to the effectiveness of the proposed surrogate modeling method, and it tends to increase in the low-probability region, as evidenced from Figure \ref{Fig:Example1_fig2}. Moreover, there is no guarantee that the heteroscedastic Gaussian process can completely capture the noisy error correction, i.e., provide an unbiased estimate of error correction. In this sense, the proposed method will inevitably exhibit certain bias for probability estimations, especially for a rare event. To address this problem, a final step of importance sampling has been introduced in this work to further improve the probability estimations. 

        }

	\section{Conclusions}\label{Sec:conclude}
	\noindent This paper develops an effective surrogate modeling method for rare event simulation with high-dimensional input uncertainties. The method circumvents the curse of dimensionality by using physics-based surrogate models parameterized by a few tunable parameters. A data-fitting error correction constructed in the output space of the physics-based surrogate model is leveraged to correct the bias of surrogate modeling. Due to inherent stochastic noise in the errors, the heteroscedastic Gaussian process is adopted to model the error correction function. An active learning process is developed to effectively train the coupled physics-data-driven surrogate model to explore the critical region for the rare event. A final importance sampling step is designed to approximate the correction factor for the surrogate model-based probability estimation. Three numerical examples are studied to demonstrate the performance of the proposed method. The first example considers a static problem of a linear elastic cantilever beam with material properties modeled by a Gaussian random field. The second example studies a dynamic problem of a nonlinear viscous damper under stochastic excitation. The third example investigates a multi-degree-of-freedom hysteretic system under stochastic excitation. Three schemes are investigated to construct physics-based surrogate models: a homogenization of material properties is considered in the first example, statistical linearization is used in the second example, and relaxation of the numerical solver is adopted in the third example. The numerical results are highly promising, suggesting that the proposed method can leverage limited calls of the original computational model to effectively estimate rare event probabilities with high-dimensional input uncertainties.

	%\section*{Acknowledgement}
	%\noindent
	%Dr. Ziqi Wang was supported by the National Science and Technology Major Project of the Ministry of Science and Technology of China (Grant No. 2016YFB0200605), National Natural Science Foundation of China (Grant No.1808149), and the Natural Science Foundation of Guangdong province (Grant No.2018A030310067). Dr. Marco Broccardo was supported by DESTESS, a projected which has received funding from the European Union's Horizon 2020 research and innovation programme under Grant No.691728. This support is gratefully acknowledged. Any opinions, findings, and conclusions expressed in this paper are those of the authors, and do not necessarily reflect the views of the sponsors.
	
	\bibliography{ReferenceList}
	
	\appendix
	\section{Implementation details}\label{Append:implementationdetails}
	
	\begin{figure}[H]
		\centering
		\includegraphics[scale=0.75]{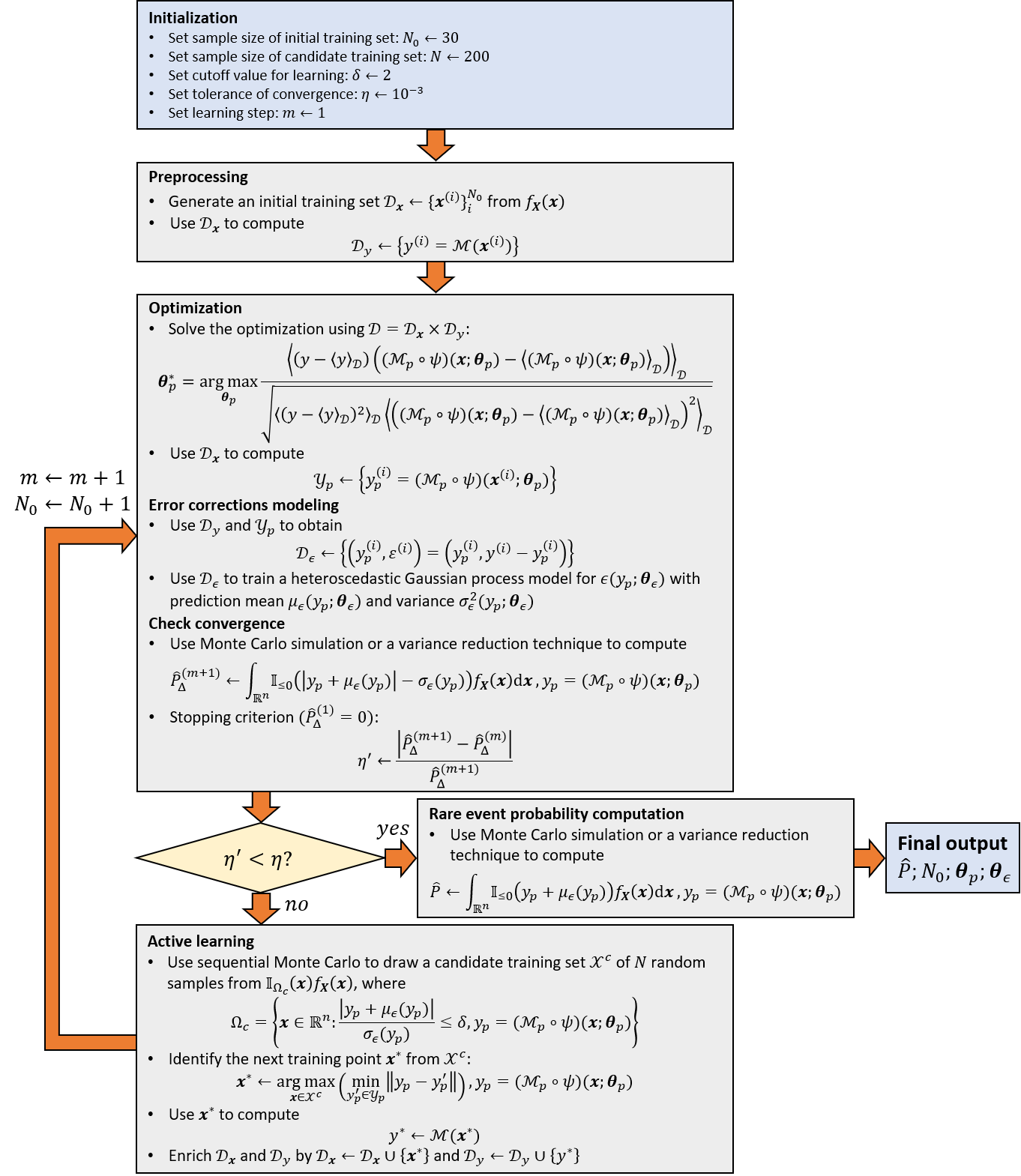}
		\caption{\textbf{{Training of the coupled physics-data-driven surrogate model}}.}
		\label{Fig:Appendix1}
	\end{figure}
	
	\begin{figure}[H]
		\centering
		\includegraphics[scale=0.82]{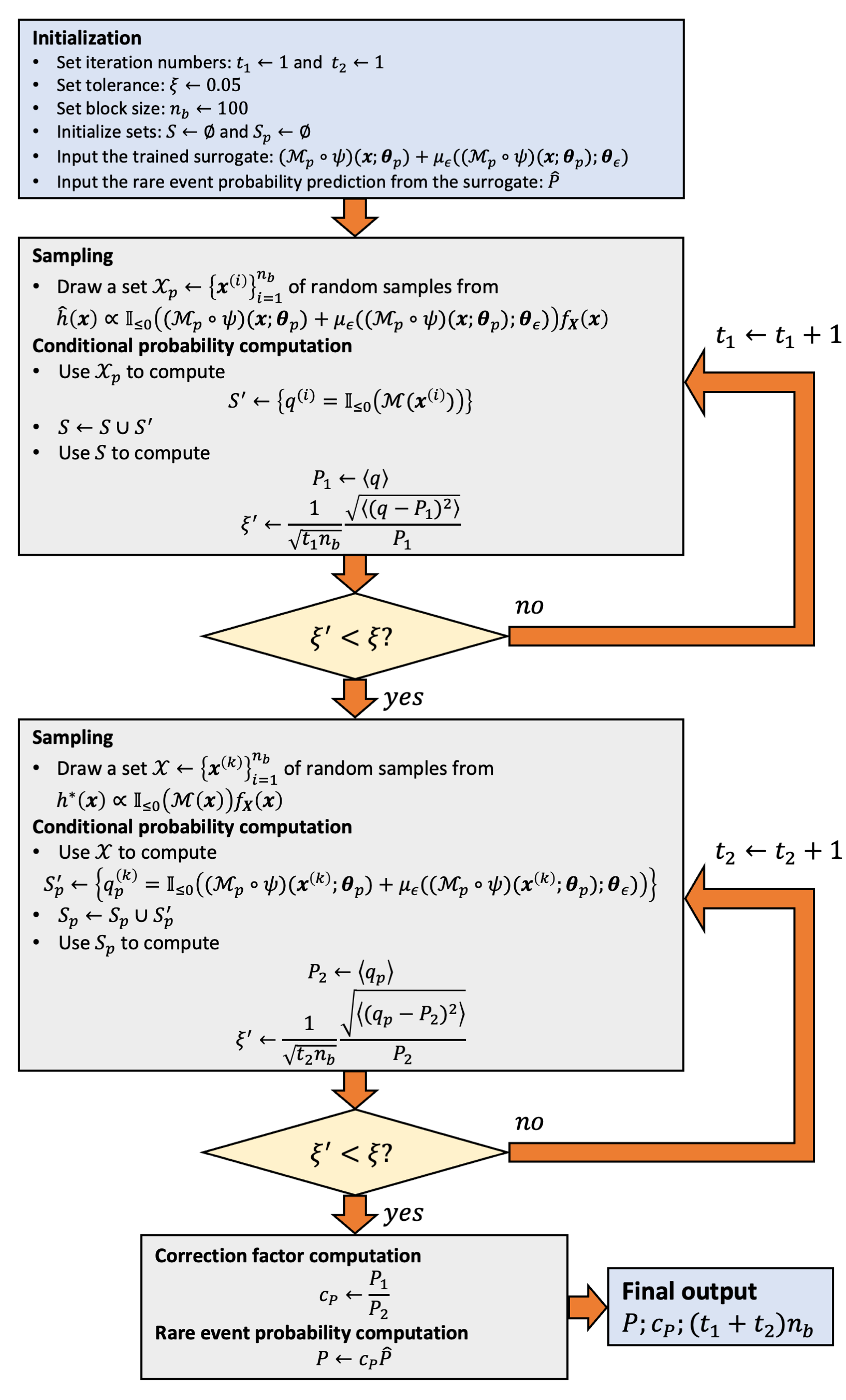}
		\caption{\textbf{Importance sampling for the coupled physics-data-driven surrogate model}.}
		\label{Fig:Appendix2}
	\end{figure}
	
\end{document}